\documentclass[11pt]{article}
\usepackage{natbib}
\usepackage{amsmath}
\usepackage{amssymb}
\usepackage[margin=1in, footskip=0.25in]{geometry}
\usepackage{amsthm}
\usepackage{placeins}
\usepackage[capposition=top]{floatrow}
\usepackage{graphicx}
\usepackage{booktabs}
\usepackage[bf]{titlesec}
\usepackage[nottoc]{tocbibind}
\usepackage{setspace}
\usepackage{lscape}
\usepackage{tikz}
\usepackage{tkz-euclide,subfigure}
\usepackage{multicol}
\usepackage{colortbl}
\usepackage{enumerate}
\usepackage{import}
\usepackage{tikz}
\usepackage[flushleft]{threeparttable}
\usepackage{longtable}
\usepackage{xtab}
\usepackage{tabulary}
\newcolumntype{K}[1]{>{\centering\arraybackslash}m{#1}}
\usepackage{threeparttable}
\usepackage{multirow}
\usepackage{fnpct}
\usepackage{pdflscape}
\usepackage{titletoc}
\usepackage[hidelinks,bookmarks=false]{hyperref}
\hypersetup{
	colorlinks,
	linkcolor={blue!80!black},
	citecolor={blue!80!black},
	urlcolor={blue!80!black},
	pdfstartview={}
}
\usepackage{booktabs}
\usepackage{caption}
\definecolor{dgray}{gray}{0.35}

\usepackage{epigraph}

\usepackage{etoolbox}

\makeatletter
\patchcmd{\epigraph}{\@epitext{#1}}{\itshape\@epitext{#1}}{}{}
\makeatother

\title{Discrimination Against Immigrants in the Criminal\\ Justice System:
Evidence from Pretrial Detentions\thanks{We thank Peter Hull, Patrick Kline, and seminar participants at the IADB-Migration group, the Online Economics of Discrimination Seminar, and UC Berkeley for very helpful discussions and suggestions. Guillermo Palacios provided outstanding research assistance. Grau acknowledges financial support from the Centre for Social Conflict and Cohesion Studies (ANID/FONDAP/15130009). Vergara acknowledges financial support from the Law, Economics, and Politics Center (LEAP) at UC Berkeley. Usual disclaimers apply.}}
\date{First version: \today\\ This version: \today}
\author{Patricio Domínguez\thanks{Department of Industrial and Systems Engineering, Pontificia Universidad Catolica de Chile, pdomingr@ing.puc.cl}
  \and
  Nicolás Grau\thanks{Department of Economics, Faculty of Economics and Business, Universidad de Chile, ngrau@fen.uchile.cl.}
  \and
  Damián Vergara\thanks{Department of Economics, UC Berkeley, damianvergara@berkeley.edu.}\\
}

\begin{document}

\maketitle

\begin{abstract}
This paper tests for discrimination against immigrant defendants in the criminal justice system in Chile using a decade of nationwide administrative records on pretrial detentions. Observational benchmark regressions show that immigrant defendants are 8.6 percentage points less likely to be released pretrial relative to Chilean defendants with similar proxies for pretrial misconduct potential. Diagnostics for omitted variable bias --including a novel test to assess the quality of the proxy vector based on comparisons of pretrial misconduct rates among released defendants-- suggest that the discrimination estimates are not driven by omitted variable bias and that, if anything, failing to fully account for differences in misconduct potential leads to an underestimation of discrimination. Our estimates suggest that discrimination stems from an informational problem because judges do not observe criminal records in origin countries, with stereotypes and taste-based discrimination playing a role in the problem's resolution. We find that discrimination is especially large for drug offenses and that discrimination increased after a recent immigration wave.

\end{abstract}

\thispagestyle{empty}

\newpage
\pagenumbering{arabic} 

\epigraph{\raggedleft ``When Mexico sends its people, they’re not sending their best. (...). They’re bringing drugs.\\ They’re bringing crime. They’re rapists. And some, I assume, are good people.'' \medskip }{\textup{Donald Trump, presidential announcement speech, June 16, 2015.}}

\vspace{-1cm}
\section{Introduction}
\label{sec1}

Increasing flows of international migrants have induced several policy challenges for destination countries in recent years \citep{un2,un} including, among others, the mitigation of concerns about discrimination \citep{heath2013discrimination}. Labor market, crime, and welfare concerns can trigger negative attitudes towards immigrants, particularly when cultural prejudice and misperceptions regarding foreign born and minority populations are commonplace \citep{dustmann2007racial,hangartner2019does,grigorieff2020does,ajzenman2021immigration,alesina2021immigration,alesinastant,bursztyn2021immigrant}. Characterizing discrimination patterns against immigrants is therefore a policy priority to improve the possibilities to integrate immigrants in destination countries.

This paper conducts an empirical analysis of discrimination against immigrant defendants in the criminal justice system, in particular, in the dictation of pretrial detention. Pretrial detention decisions are particularly relevant for two main reasons. First, one of the prevalent narratives that underpins negative attitudes towards immigration is its potential effect on crime \citep{fasani2019does,bursztyn2021disguising}. Thus, if there is discrimination against immigrants, it should be especially salient in institutions that are concerned with that rubric. Second, the pretrial detention process has been shown to be permeated by discriminatory practices \citep{ady,arnold2020measuring,pbot} and to have a significant impact on the labor market opportunities, conviction probabilities, and the participation in public assistance programs for defendants \citep{heaton2017downstream,leslie2017unintended,dgy,dobbie2021economic,dobbie2021jep,grau2019effect,stevenson2021pretrial}. These costs are particularly harmful for immigrant populations given the challenges they face relating to integration in general and entry into the labor market in particular \citep{aaslund2014seeking,dustmann2017economics,brell2020labor,dennis}. 

We study this scenario in Chile where the pretrial detention process functions in a similar manner to the US---that is, judges decide on the pretrial detention status of defendants using limited information and in a limited time period with a mandate to not detain defendants unless misconduct potential (pretrial recidivism or failure to appear in court, or both) is considerable. Unlike in the US, however, the Chilean system is uniform across all localities and does not permit monetary bail. Since approximately 2013 Chile has become an important destination for immigrants from other Latin American countries. And, as we discuss in Section \ref{sec3}, this immigration flow has generated hostile discourses in the public sphere and misperceptions regarding immigrants and immigration in the general population. 

We conduct the empirical analysis using nationwide administrative records for the period 2008---2017. Our database covers 95\% of Chile's criminal prosecutions and includes detailed information on defendants and prosecutions. Despite having cleaner criminal records upon prosecution and lower pretrial misconduct rates when released, immigrant defendants are more likely to be detained pretrial relative to defendants who are Chilean nationals. We formally explore whether this descriptive fact constitutes some form of aggregate discrimination by estimating observational benchmark regressions---that is, by testing if defendants with similar proxies for misconduct potential face, on average, different pretrial release rates because of their immigration status, an exercise similar in spirit to \cite{gelman2007analysis}, \cite{abrams2012judges}, and \cite{rehavi2014racial}. The main result shows that, conditional on type of crime, criminal records, judge leniency, attorney quality, and court-by-time fixed effects, immigrant defendants are 8.6 percentage points (pp) more likely to be detained pretrial relative to Chilean defendants, a difference that represents 53\% of the unconditional baseline pretrial detention rate. 

We perform two sets of diagnostics that suggest that our main estimates are not driven by systematic differences in unobserved misconduct potential (i.e., omitted variable bias, OVB). First, we implement the method set out in \cite{oster2019unobservable} for assessing the degree of selection on unobservables using observed covariates. We conclude that, if anything, our observational benchmark regression underestimates discrimination against immigrants. Specifically, we show that the discrimination estimate is attenuated (although not reversed) only if we impose the nonstandard assumption that unobservables are negatively correlated with the control variables---that is, if defendants with observables that induce lower release rates are also endowed with unobservables that work in the opposite direction. This is consistent with the fact that immigrants have, on average, cleaner criminal records. In fact, and contrary to the common finding in observational benchmark regressions, we find that the release disparity between Chilean defendants and immigrant defendants is 30\% smaller (6.6 pp) when control variables are omitted.

The second diagnostic is a novel test that uses observed pretrial misconduct rates of released defendants to assess the degree of OVB in the release equation. The intuition is as follows: OVB arises if there are unobservables that correlate with immigration status and predict misconduct potential. Then, we can use observed pretrial misconduct of released defendants to test whether the included control variables are sufficient to account for systematic differences in pretrial misconduct rates between immigrant defendants and Chilean defendants. Because the sample of released defendants is selected, we estimate different selection models using assigned judges and attorneys as the excluded variables and show that, conditional on type of crime, criminal records, and time-varying court characteristics, immigration status has no predictive power on observed pretrial misconduct. This confirms that the set of controls included in the benchmark regression is sufficient to account for systematic risk differences between Chilean defendants and immigrant defendants. 

The second test also suggests that, while time-varying court characteristics matter for the discrimination point estimate, they do little to correct for OVB and, therefore, they may induce included variable bias (IVB) in the main estimations. In other words, rather than controlling for differences in unobserved misconduct potential, court-by-time variation seems to mediate discrimination through the systematic assignment of immigrant defendants to less lenient courts. Under this interpretation, the baseline discrimination estimate of 8.6 pp constitutes a conservative bound given that excluding the court-by-time fixed effects increases the release disparity gap to 10.3 pp. 

After establishing the robustness of the reduced-form discrimination estimate, we perform different heterogeneity analyses to obtain additional insights about the aggregate pattern. We first explore whether discrimination against immigrants stems from an informational problem. The criminal records of immigrant defendants are likely to be censored because judges cannot observe the records kept in the country of origin. Then, judges may process the information contained in criminal records differently depending on whether the defendant is immigrant or not. To test this hypothesis, we estimate our main specification separately for samples of defendants with and samples of defendants without previous prosecutions in Chile. We find that the conditional release disparities are 1.9 pp and 13.2 pp for defendants with and defendants without previous prosecutions, respectively. This implies that, unlike Chilean defendants, immigrant defendants are not rewarded for having clean criminal records. In fact, comparing the discrimination estimates with the unconditional differences in release rates suggest that judges systematically impute nonzero criminal histories for all immigrant defendants, regardless of their criminal records in Chile. We complement this analysis by showing that the discrimination estimate is 36\% smaller for experienced judges. Both pieces of evidence strongly suggest that the informational problem seems to be the main driver behind the discrimination patterns. 

The fact that discrimination stems from an informational problem does not necessarily mean that the aggregate estimate can be rationalized by accurate statistical discrimination. On one hand, the informational problem can be circumvented based on stereotypes (biased beliefs). On the other hand, the informational problem may provide the room for judges to exercise preferences against immigrants (taste-based discrimination). To obtain further insights, we estimate observational outcome tests \citep{kpt,pbot} that, when rejected, provide evidence of a nonseparately identified combination of stereotypes and taste-based discrimination \citep{hull2021}. The outcome test is rejected suggesting that the discrimination pattern, although stemming from an informational problem, cannot be rationalized by accurate statistical discrimination. The combination of stereotypes and taste-based discrimination seems to play an important role in explaining the aggregate estimates of discrimination against immigrants.

We also explore whether or not the patterns of discrimination vary according to crime category. We find that discrimination is particularly severe for drug offenses: conditional release disparities are around four times larger than the baseline benchmark estimate. Statistically and economically significant differences are also found for thefts and robberies and other property crimes. Other violent crimes like homicides and sexual offenses show no significant differences, although the number of immigrants that are charged with these types of crimes is too small to confidently detect an effect. Crime categories that are associated with less hostile discourses in the public sphere (such as white collar crimes or crimes against privacy) also display no significant release disparities. This heterogeneity is consistent with the discourse that links immigration flows to drug trafficking and other crimes that tend to be prominent in surveys that measure citizens perceived fears of crime playing a role in affecting preferences and stereotypes.

Finally, we explore the evolution of discrimination patterns during the recent wave of immigration to Chile. Immigration inflows in Chile grew considerably and became more racially diverse after 2012. We find that benchmark estimates are 55\% larger in the 2013--2017 period relative to the 2008--2012 period. We also find that discrimination estimates are remarkably similar across different groups of immigrants, suggesting that the increase over time is driven by an intensification of discrimination rather than a composition effect. Although we do not establish a causal relationship, we conjecture that this trend is related to the crime-related narrative response to the immigration wave lead by political authorities and the media \citep{ajzenman2021immigration}.

\section{Setting and Data}
\label{sec3}

This section describes the Chilean context and institutional setting. We start by characterizing immigration in Chile, discussing its magnitude, trends, composition, and the attitudes to immigration among Chilean citizens. We then describe the pretrial detention system. Finally, we describe the data used in the empirical analysis and provide descriptive statistics.  

\subsection{Immigration in Chile}

Immigration in Chile has changed dramatically since 2013. According to the national census, immigrants represented around 1.5\% of the Chilean population in 2002 and in 2012 this figure was 2\%. In 2017, however, the share of immigrants in the population increased to 4.5\%, and projections from the National Institute of Statistics suggest that the number reached 7.8\% in 2019. In terms of gross inflows, both the volume and composition has changed considerably. Between 2008 and 2012 the inflow of immigrants was fairly stable at around 100,000 individuals per year. Peruvians accounted for the largest share of immigrants arriving each year (roughly 50\% of new arrivals) and Argentinians, Bolivians, and Colombians collectively represented around 25\% of inflows. The annual inflow of immigrants grew to around 350,000 individuals in 2017. In addition, the composition changed substantially. For the 2008--2012 period, Haitian and Venezuelan immigrants accounted for around 2\% of the arrivals; however, for the 2013--2017 period they represented almost 20\% of the arrivals.\footnote{To illustrate the prominence of Chile as an destination of choice within the region, in March 2021 Chile was the country with the second highest number of Venezuelan immigrants equating to 600,000 individuals with regular status, just below Colombia with 720,000. See \href{https://data2.unhcr.org/en/situations/vensit}{https://data2.unhcr.org/en/situations/vensit} for details.} Inflows of Colombian nationals also increased significantly after 2013. Their share among new arrivals to Chile grew from 10\% in the period 2008--2012 to 19\% in the period 2013--2017.

The recent immigration wave has not gone unnoticed by Chile’s population, politicians, and media. Similar to the inhabitants of other countries \citep{hopkins2019muted,jorgensen2020correcting,alesinastant}, Chileans overestimate the size of the immigrant population. As summarized in \cite{ajzenman2021immigration}, surveys show that in 2019 Chileans thought that immigrants represented 33\% of the total population, which compares unfavorably to the administrative estimate of 7.8\%. Additional survey evidence shows that \textit{increase in crime} is the most prominent concern related to immigration among Chile’s population. \cite{ajzenman2021immigration} report that these perceptions are not related to changes in victimization rates, even though the concerns we severe enough to induce increases in preventive behavioral responses, such as investments in home security. In addition, \cite{ajzenman2021immigration} show that the media plays an important role in the divergence between perceptions and actual crime rates, for example, by increasing the number of crime-related news segments with an alleged perpetrator that is an immigrant.

Public attitudes toward immigrants can also be inferred from recent political campaigns. Immigration, which was not a public issue before the recent increase in immigration inflows, has become a prominent topic in the national political debate, with an increasing number of politicians framing their public discourse in the context of immigration regulation. For example, inspired by Donald Trump, the far-right candidate in the last presidential election, Jos\'e Kast, proposed to build a ditch on Chile's border with Bolivia. In most cases, a link between crime and immigration is proffered as the most important reason for stopping or reducing the arrival of new immigrants. For example, when the National Congress of Chile passed President Piñera's migration project in 2019, he described the initiative as a ``(...) great step forward to continue putting our house in order regarding matters of migration, and thus better combat illegal immigration and the evils such as delinquency, drug trafficking, and organized crime entering Chile'' (Twitter, January 16, 2019). These types of statements, that display either implicit or explicit anti-immigrant sentiment, have been accompanied by instances of misinformation that appear to be designed to stigmatize immigrants or immigration. For example, in 2018, a widely shared social media post claimed that the former President Bachelet received USD 3,000 from the UN for each Haitian immigrant that entered Chile. The increase in negative attitudes towards immigrants has contributed towards Chile's political climate becoming more polarized and an increase in the relative popularity of far-right nationalist groups with conservative-leaning voters.\footnote{There is a contrast between the hostile expressions against immigrants, such as President's Piñera's tweet, with what is usually presented as the traditional view on Chilean attitudes towards immigration as expressed in Chito Faró’s famous 1942 folk song ``Si vas para Chile'' (``If you go to Chile''). This popular and traditional song describes Chileans as particularly receptive to foreigners stating: ``Y verás como quieren en Chile al amigo cuando es forastero'' (``and you will see how they love a friend in Chile when is a foreigner''). Other music artists have offered a different perspective on attitudes towards immigration. For example, a 2012 track by Chilean rapper Portavoz called ``Poblador del mundo'' (``World citizen'') tells a story in the voice of an immigrant that feels constant discrimination living in Chile. In one verse, he raps ``Llegué con fe para doblarle la mano al destino y me encontré con que aquí soy un enemigo. Y es que con mi llegada los dramas también han venido, ya me han hecho entender que no soy para nada bienvenido'' (``I arrived with the belief that I could turn the hand of destiny but found that here I am an enemy. With my arrival the dramas have also come, they have already made me understand that I am not welcome at all'').}

\subsection{Pretrial Detentions in Chile}

The procedure to define pretrial detention for people arrested in Chile works as follows: During the 24 hours after the initial detention, there is an arraignment hearing in which a detention judge must decide whether the defendant will be incarcerated during the criminal investigation. Monetary bail is not an option in the Chilean system. Following the legal principle of the presumption of innocence, judges should not incarcerate defendants unless there is clear danger of escape (i.e., a high probability of failing to appear in court), the defendant represents a significant threat to society (i.e., a high probability of committing another crime during the investigation), or the imprisonment of the defendant aids the investigation of the criminal case. As described in \cite{grau2019effect}, pretrial detention became more frequent between 2007 (17,891 cases) and 2018 (34,815 cases).  In the same period, the proportion of cases where pretrial detention is imposed has increased from 7.3\% to 9.6\%, and pretrial detainees as a share of total prisoners rose from 21.9\% to 36\%.

Pretrial detention hearings are brief (lasting about 15 minutes) and judges have to make a decision whether or not to detain the defendant at the end of the hearing. Judges are assigned in a quasi-random fashion at the court-by-time level: in every court at the beginning of each month judges are assigned to different time slots to lead arraignment hearings for no reason other than splitting this duty among the judges in the particular court. The assignment of public attorneys to each case is also quasi-random at the court-by-time level.

\subsection{Data}

We use administrative records from the Public Defender's Office (PDO). The PDO is a centralized public service under the oversight of the Ministry of Justice. It offers criminal defense services to all individuals accused of or charged with a crime; as such, it ensures the right to a defense by a lawyer and due process in criminal trials. Our estimation sample covers more than 95\% of the criminal cases for the period between 2008 and 2017, and contains detailed case and defendant characteristics, including the nationalities of the defendants. We define immigrant defendants as defendants that are nationals of any other country than Chile. In addition, we can identify the judges assigned to each case at the beginning of the criminal process (i.e., when pretrial detention decisions take place), as well as the public attorney assigned to the arraignment hearing.
 
To build the estimation sample, we consider all detention hearings for adult defendants who were arrested between 2008 and 2017 and defended by the PDO. We exclude hearings due to legal summons because the information set available to the judge may be different in those cases. To be able to focus on arraignment hearings in which pretrial detention is a plausible outcome, we only consider types of crimes that have at least a 5\% probability of pretrial detention. For the same reason, when defendants are accused of more than one crime during the same arraignment hearing, we only retain the information related to the most severe crime (with severity measured as the probability of pretrial detention). A more detailed description of the data, the sample restrictions, and the variables is presented in Appendix \ref{data_app}.

\begin{table}[t!]
\RawCaption{\caption{Descriptive Statistics}\label{Tab_Desc_est}}
{}
{\small\begin{tabular}{lcc}\toprule
 & Chilean & Immigrant   \\ \cline{2-3}
 &  &     \\ [-3pt]
 Released                                    &         0.84 &         0.77   \\ [5pt]
 \textbf{Outcomes (only for released)}       &  &      \\ [3pt]
 Nonappearance in court                     &         0.17  &         0.16   \\ [3pt]
 Pretrial recidivism                         &         0.19  &         0.13   \\ [3pt]
 Pretrial misconduct                         &         0.29  &         0.24   \\ [5pt]
 \textbf{Individual Characteristics}         &  &       \\ [3pt]
 Male                                        &         0.88  &         0.87   \\ [3pt]
 At least one previous case                  &         0.68  &         0.41   \\ [3pt]
 At least one previous pretrial              &         0.40  &         0.22   \\ [3pt]
 misconduct                                  &  &       \\ [3pt]
 At least one previous conviction            &         0.65  &         0.39   \\ [3pt]
 No. of previous cases                       &         4.58  &         2.48   \\ [3pt]
 Severity previous case                      &         0.09  &         0.05   \\ [3pt]
 Severity current case                       &         0.18  &         0.18   \\ [5pt]
 \textbf{Court Characteristics}  &  &      \\ [3pt]
 Average severity (year/Court)               &         0.09 &         0.11   \\ [3pt]
 No. of cases  (year/Court)                  &        3,023 &        3,557   \\ [5pt]
 No. of judges (year/Court)                  &           46 &           58   \\ [5pt]
 \textbf{Observations (released)} &      580,400 &        4,900  \\
 \textbf{Observations (nonreleased)} &      112,606 &        1,462  \\
\bottomrule
\end{tabular}
\captionsetup{justification=justified,margin=0cm}
\floatfoot{\footnotesize\textbf{Notes:} This table presents descriptive statistics for our estimation sample. The sample considers all arraignment hearings for adult defendants who were arrested between 2008 and 2017. We do not include hearings due to legal summons and only consider types of crimes with at least a 5\% probability of pretrial detention. When defendants are accused of more than one crime, we retain the information related to the most severe crime (with severity measured as the probability of pretrial detention). More details about the data and the variables can be found in Appendix \ref{data_app}.}}
\end{table}

Table \ref{Tab_Desc_est} shows descriptive statistics for the estimation sample. This table provides three stylized facts that motivate our research question. First, relative to Chilean defendants, immigrants are more likely to be detained before the trial. The pretrial release rate for immigrant defendants and Chilean defendants is 77\% and 84\%, respectively. Second, immigrant defendants have cleaner criminal records. Chilean defendants are more likely to have previous prosecutions, convictions, and instances of pretrial misconduct. The criminal records of immigrants may be censored because judges cannot observe the criminal histories of immigrant defendants before they arrive to Chile. Therefore, how judges interpret immigrants' (lack of) criminal records could play a role in explaining the release disparities. Third, once released, immigrants are less likely to be engaged in any kind of pretrial misconduct; pretrial recidivism rates are 6 pp larger for Chilean defendants. 

\section{Theoretical Framework}
\label{sec2}

To motivate a formal test for discrimination in pretrial detention decisions, this section develops a simple theoretical framework to guide the empirical analysis and formally defines the notion of discrimination we use throughout the paper.

\subsection{Model}

Let $i$ index defendants and $j$ index judges, with judges assigned to defendants according to the mapping $j(i)$. Let $Y_i^*\in\{0,1\}$ be pretrial misconduct if released, with $Y_i$ being the realized value, so $Y_i=Y_i^*$ for pretrial released defendants and $Y_i = 0$ for pretrial detained defendants. Judges are mandated to detain defendants with $Y_i^*=1$ and release defendants with $Y_i^*=0$ but they do not observe $Y_i^*$ at the time of the decision.\footnote{Alternatively, $Y_i^*$ could be defined as the probability of defendants engaging in some form of pretrial misconduct if released, with $Y_i^*\in[0,1]$ and the judges' mandate to detain defendants pretrial when $Y_i^*$ exceeds some threshold.} Instead, judges base the release decision on their predictions on the probability that $Y_i^*=1$. The prediction is made using all the available information, which includes the nationality of the defendant, $I_i=1\{\text{$i$ is immigrant}\}$, and a vector of other characteristics, $Z_i$. 

The principle of the presumption of innocence and the documented impact of pretrial detention imply that only defendants with a high predicted probability of pretrial misconduct should be detained. Because the decision is discretionary and judges potentially have heterogeneous preferences, release thresholds are idiosyncratic. Then, the release equation for defendant $i$ can be written as
\begin{eqnarray}
Release_{i} &=& 1\left\{p_{j(i)}(I_i,Z_i) \leq t_{j(i)}(I_i,Z_i)\right\},\label{release}
\end{eqnarray}
where $p_{j(i)}(I_i,Z_i)\in[0,1]$ is the prediction of the pretrial misconduct probability of defendant $i$  made by the assigned judge $j(i)$, and $t_{j(i)}(I_i,Z_i)\in[0,1]$ is the judge-specific release threshold. 

Let $p(I_i,Z_i) = \Pr(Y_i^*=1|I_i,Z_i)$. Because the conditional distribution is possibly unknown to judges, their predictions regarding pretrial misconduct are potentially biased. Let $b_{j(i)}(I_i,Z_i)$ be the judge-specific bias in predictions, such that $p_{j(i)}(I_i,Z_i) = p(I_i,Z_i) + b_{j(i)}(I_i,Z_i)$. Then,
\begin{eqnarray}
Release_{i} = 1\left\{p(I_i,Z_i) \leq t_{j(i)}(I_i,Z_i) - b_{j(i)}(I_i,Z_i)\right\} \equiv 1\left\{f_{j(i)}(I_i,Z_i) \geq 0\right\} = R(I_i,Z_i,j(i)). \label{release2}
\end{eqnarray}
As \cite{hull2021} stresses, this analysis does not require additional assumptions on judge behavior, relating to the way they set thresholds or make predictions, for example. The analysis only needs the judge decision to be binary and to follow a threshold-crossing model. As discussed in Section \ref{sec3}, both assumptions represent a good approximation of the Chilean setting. In this simple model, the decision is deterministic from the judge's perspective. We denote the function that determines whether a defendant with immigration status $I_i$ and characteristics $Z_i$ is released or not when assigned to judge $j(i)$ by $R(I_i,Z_i,j(i))\in\{0,1\}$.

\subsection{Sources of Release Disparities and Normative Definitions}

Equation \eqref{release2} suggests that a defendant's nationality may affect the release decision through different channels. First, judges may use $I_i$ to predict (or interpret signals of) variables they do not observe that affect $Y_i^*$ if distributions vary with $I_i$. This is defined as statistical discrimination \citep{aigner1977statistical} and directly affects the conditional outcome probabilities, $p(I_i,Z_i)$, and the bias, $b_{j(i)}(I_i,Z_i)$, if these parameters impact how priors are updated. Second, if statistical discrimination is inaccurate and the degree of inaccuracy varies with $I_i$, this induces a differential prediction bias to the decision. We refer to this as biased beliefs or stereotypes, and its effect is captured by $b_{j(i)}(I_i,Z_i)$ \citep{bordalo2016stereotypes,bohren2020inaccurate}. Third, release thresholds may depend on $I_i$ because of animus. This affects $t_{j(i)}(I_i,Z_i)$ and gives form to the standard notion of taste-based discrimination or discriminatory preferences \citep{b1,b2}. 

In this paper, we focus on an aggregate notion of discrimination that incorporates the three sources of release disparities mentioned: statistical discrimination, biased beliefs or stereotypes, and taste-based discrimination. A possibly complex combination of these drivers of release disparities may yield a reduced-form effect of $I_i$ on $Release_i$, captured by the function $f_{j(i)}(I_i,Z_i)$ even within defendants with equal $Y_i^*$. As emphasized in \cite{yang2020equal} and \cite{arnold2020measuring}, all sources of disparities are problematic through the lens of legal principles similar to the Equal Protection Clause in the US Constitution, which suggests that this reduced form effect has normative content regardless of the specific sources. Intuitively, the composite effect of all sources may lead defendants with equal misconduct potential but different immigration status to face different pretrial detention rates. This motivates the following definition.

\noindent\textsc{Definition (Discrimination)}: \textit{discrimination, $\mathbf{D}$, is defined as}
\begin{eqnarray}
\mathbf{D} = \mathbb{E}[\mathbf{d}(Y_i^*)] = \mathbb{E}[\mathbb{E}[d(Z_i,j(i))|Y_i^*]] = \mathbb{E}[\mathbb{E}\left[R(1,Z_i,j(i)) - R(0,Z_i,j(i))|Y_i^*\right]], \label{TD_1}
\end{eqnarray}
\textit{where the inner expectation integrates over $(Z_i,j(i))$, and the outer expectation integrates over $Y_i^*$.}

Under this discrimination definition, there is system-wide discrimination against immigrants if $\mathbf{D}<0$. The inner object, $d(Z_i,j(i))$, is the judge-specific gap in release rates between immigrant and native defendants endowed with $Z_i$. $\mathbf{d}(Y_i^*)$ is the expectation of the judge-specific measure over $(Z_i,j(i))$ conditional on misconduct potential $Y_i^*$ and, therefore, does not impose restrictions on the group-specific distributions. Finally, the aggregate measure of discrimination averages over the distribution of misconduct potential. Similar to the definition outlined in \cite{arnold2020measuring}, this definition compares release rates between immigrant and native defendants with equal pretrial misconduct potential, with potential differences arising because of statistical discrimination, biased beliefs, or taste-based discrimination, or a combination of these sources of discrimination. 

\paragraph{Empirical test} One fundamental challenge when testing for discrimination following \eqref{TD_1} is that $Y_i^*$ is most likely not observable; therefore, a regression of $Release_i$ on $I_i$ controlling for $Y_i^*$ is potentially not implementable. \cite{arnold2020measuring} propose a weighting scheme that equalizes group-specific distributions of $Y_i^*$ under which omissions of $Y_i^*$ from the (weighted) regression of $Release_i$ on $I_i$ do not induce OVB and, therefore, identifies $\textbf{D}$. Unfortunately, the immigrant population in Chile is small, which prevents us from implementing this methodology.\footnote{The \cite{arnold2020measuring} method combines quasi-random assignment of judges, observational misconduct data on released defendants behavior, and extrapolation techniques. The extrapolation step requires several judges handling several cases that involve defendants that are immigrants. Our limitation is, therefore, statistical power. Specifically, the extrapolation technique used in \cite{arnold2020measuring} is based on regressing average pretrial misconduct rates of released immigrant defendants against judge leniency to infer the average pretrial misconduct rate of a supremely lenient judge using an identification at infinity argument. For this technique to work properly several judges are required in order to have enough support, in addition to several cases per judge so that pretrial misconduct rates can be precisely estimated. Both requirements are not met in our immigrant subsample.} As an alternative, we implement observational benchmark regressions similar in spirit to \cite{gelman2007analysis}, \cite{abrams2012judges}, and \cite{rehavi2014racial}. Formally, we assume that $Y_i^*$ can be expressed as a function of observed variables, $X_i$, so the discrimination definition can be written as
\begin{eqnarray}
\mathbf{D} = \mathbb{E}[\mathbf{d}(g(X_i))] = \mathbb{E}[\mathbb{E}[d(Z_i,j(i))|g(X_i)]] = \mathbb{E}[\mathbb{E}\left[R(1,Z_i,j(i)) - R(0,Z_i,j(i))|g(X_i)\right]], \label{TD}
\end{eqnarray}
where $Y_i^*=g(X_i)$. Although $X_i$ and $Z_i$ possibly overlap, $X_i$ does not have to be equal to $Z_i$. Under this assumption, regressing $Release_i$ on $I_i$ controlling for $g(X_i)$ identifies $\textbf{D}$.

In Section \ref{sec5} we discuss our choice of $X_i$ and provide diagnostics to assess its performance in approximating $Y_i^*$. This is central to the analysis because the extent to which \eqref{TD} constitutes a good approximation of \eqref{TD_1} fundamentally depends on the vector $X_i$. More specifically, as noted by \cite{arnold2020measuring}, observational benchmark regressions can suffer both from IVB and OVB. IVB arises when some of the included variables in $X_i$ mediate discrimination against, in this case, immigrant defendants in a disparate impact sense. For example, if there is discrimination based on place of living and immigrants are overrepresented in a discriminated locality, then controlling for place of living will attenuate the estimated disparity. OVB arises when the vector $X_i$ does not fully account for group differences in misconduct potential and, therefore, the estimated coefficient reflects differences in unobservables rather than discriminatory forces. This empirical challenge generates a tension between over and under controlling. 

We deal with this tension by choosing the smallest set of variables that successfully eliminates OVB and therefore minimizes the likelihood of IVB.  Section \ref{sec5} presents several tests that reject OVB given our choice of $X_i$ and develops sensitivity analyses that flag potential sources of IVB in our setting. In particular, variables that do little to attenuate OVB concerns and significantly affect the discrimination estimate are more likely to contaminate the result with IVB.

\paragraph{On the normative interpretation} Related to the discussion on IVB and OVB, our definition of discrimination seeks to identify a disparate impact (rather than disparate treatment) notion of discrimination \citep{arnold2020measuring,rose2022constructivist}. The disparate impact doctrine focuses on release disparities among defendants with equal misconduct potential but different group membership. Under this perspective, different structural models that rationalize similar release gaps conditional on misconduct potential are normatively equivalent. On the other hand, the disparate treatment doctrine focus on the actual release decision, understanding as discrimination explicit judge-level practices against a particular group. This implies that the normative implication of release disparities among defendants with equal misconduct potential is different if they follow explicit discriminatory practices against group membership or if, for example, are driven by discriminatory practices against other characteristics that correlate with group membership.

As before, the extent to which our definition relates to each normative perspective depends on the variables included in $X_i$. More specifically, it is dependent on how accurately these variables proxy misconduct potential. If $X_i$ exactly matches the information set of judges ($Z_i$), the definition will probably approximate a disparate treatment notion of discrimination because judges observe variables through which discrimination towards immigrants manifests itself and that are unlikely to predict misconduct potential on top of other controls. That is why we purposely exclude (and why we are not concerned that we do not observe) all demographics other than immigration status that are potentially observed by judges (age, place of living, physical aspects, or income, for example). By contrast, we include the minimum set of variables that, we argue, are sufficient to eliminate OVB because they constitute a reasonable proxy of misconduct potential---namely, criminal records and case characteristics.\footnote{\cite{arnold2020measuring} show that discrimination estimates are biased when criminal record and type of crime variables are included in regressions that already control for misconduct potential through an IVB mechanism, which implies that these variables are imperfect proxies for misconduct potential. This can be explained if, for example, the distributions of types of crime differ by immigration status and judges are engaged in crime-specific discrimination patterns, or if there is discrimination against immigrants in previous stages of the judicial process that is manifest in differences in the criminal records. Note, however, that \cite{arnold2020measuring} estimated observational release disparities that control for type of crime and criminal records are close in magnitude to the weighted benchmark regressions that recover the unbiased estimate, which suggests that this set of controls do a reasonably good job when approximating potential misconduct. They document unconditional gaps of 7.2 pp that are reduced to 5.2 pp when including controls and fixed effects. The bias-corrected weighted regressions with no controls report discrimination estimates that range from 4.2 pp to 5.4 pp, depending on the specific extrapolation technique.}

\section{Empirical Analysis}
\label{sec5}

This section presents the empirical strategy and the main results. We estimate substantial discrimination against immigrant defendants in pretrial detention decisions. We discuss and implement different tests for assessing the degree of OVB and conclude that unobservables are unlikely to explain the results. If anything, unobserved differences in misconduct potential are likely to exert downward bias on the discrimination estimates.

\subsection{Benchmark Regressions}

\paragraph{Empirical strategy} Discrimination, as defined by equation \eqref{TD}, can be potentially identified by \textit{benchmark regressions}---that is, by reduced-form regressions of $Release_i$ on $I_i$ and $g(X_i)$. We assume that $g$ is linear in $X_i$---that is, $g(X_i) = X_i'\alpha_X$.  And, importantly for the discussion on OVB, we further define $X_i = [X_i^o\quad X_i^u]$, where $o$ accounts for variables that are observed by the econometrician and $u$ accounts for variables that are unobserved by the econometrician. The observational benchmark test therefore consists of regressions of the following type:
\begin{eqnarray}
Release_i &=& \alpha_0 + \alpha_D I_i + X_i^{o\prime}\alpha_{Xo} + \varepsilon_i,\label{benchmark}
\end{eqnarray}
where $\varepsilon_i = X_i^{u\prime}\alpha_{Xu} + \xi_i$ is the error term. We distinguish between two sources of unobserved heterogeneity that affect the release decision. First, $X_i^{u\prime}\alpha_{Xu}$ accounts for unobservables that predict misconduct potential. The vector $X_i$ that accurately predicts $Y_i^*$ may be extensive and may therefore contain information that is not observed by the econometrician. Second, $\xi_i$ accounts for unobservables that do not predict misconduct potential but may affect the release decision. For example, $\xi_i$ may contain variables that judges use to discriminate on top of immigration status but that are not correlated with $Y_i^*$ after controlling for $X_i$. 

Let $\widehat{\alpha}_D$ be the ordinary least squares (OLS) estimator of \eqref{benchmark}. The main identification assumption for $\mathbb{E}\left[\widehat{\alpha}_D|I_i,X_i^o\right] = \mathbf{D}$ is $\mathbb{C}(I_i,X_i^u) = 0$. That is to say, the unobservables that predict misconduct potential, $X_i^u$, are not correlated with immigration status, $I_i$. This follows from equation \eqref{TD} because discrimination is defined as average release disparities conditional on misconduct potential. If $I_i$ and $X_i^u$ are correlated, then $\widehat{\alpha}_D$ will capture both discrimination and unobserved differences in misconduct potential and will therefore be contaminated by standard OVB. Note, however, that correlation between $I_i$ and $\xi_i$ is allowed by our definition of discrimination given our focus on a disparate impact notion of discrimination so its presence is, therefore, not problematic for the identification of $\mathbf{D}$. After presenting the main results, we discuss and implement diagnostics to assess the pervasiveness of potential OVB in our main benchmark estimations. 

We estimate equation \eqref{benchmark} using OLS and include the following sets of controls in the vector $X_i^o$. First, we include individual-level characteristics related to the criminal history and the type of crime. These include an indicator for whether the individual has previous prosecutions, the number of previous prosecutions, the severity of the last prosecution (measured as the average pretrial detention rate of the type of crime), an indicator for whether the individual was engaged in pretrial misconduct during a previous prosecution, an indicator for whether the individual has been convicted of a crime in the past, and type of crime fixed effects for the current prosecution. These variables are expected to proxy individual-level misconduct potential. Second, to improve precision, we include measures of judge leniency and public attorney quality and their squares, measured as residualized (against court-by-year fixed effects) leave out release rates, as in \cite{dgy}. Third, we include court-by-year fixed effects that can potentially account for unobserved shocks that correlate with misconduct potential. Standard errors are clustered at the court-by-year level. 

\begin{table}[t!]
\captionsetup{justification=centering,margin=0cm}
\RawCaption{\caption{Main Results--Benchmark Regressions: $\widehat{\alpha}_D$}\label{Tab_BT_main}}
{}
{\small\begin{tabular}{lccccc}\toprule
 & (1) & (2) & (3) & (4) & (5)   \\ \cline{2-6}
 &  &  &  & &   \\ [-1pt]
 Immigrant                                   &       -0.067       &       -0.103       &       -0.063       &       -0.030       &       -0.087        \\ [2pt]
                                             & (0.014)    & (0.013)    & (0.013)    & (0.012)    & (0.012)\\ [5pt]
 Mean dep. variable                          &         0.84 &         0.84 &         0.84 &         0.84 &          0.84 \\ [5pt]
 Individual controls                         & No & Yes & No  & No  & Yes \\ [3pt]
 Attorney and judge controls                 & No & No  & Yes & No  & Yes \\ [3pt]
 Court-by-year fixed effects                 & No & No  & No  & Yes & Yes \\ [3pt]
 No. of Immigrants     &        6,362  &        6,362  &        6,362   &        6,362  &        6,362   \\ [2pt]
 No. of Chileans       &      693,006 &      693,006 &      693,006  &      693,006 &      693,006  \\ [2pt]
 R-squared             &        0.000   &        0.167   &        0.011    &        0.023   &        0.191    \\ [2pt]
\bottomrule
\end{tabular}
\captionsetup{justification=justified,margin=0cm}
\floatfoot{\footnotesize\textbf{Notes:} This table presents the estimated benchmark coefficient, $\widehat{\alpha}_D$, of equation \eqref{benchmark} and its corresponding standard error in parentheses (clustered at the court-by-year level). Each column represents a different regression that includes different sets of controls. \textit{Individual controls} include an indicator for whether the individual has previous prosecutions, the number of previous prosecutions, the severity of the last prosecution (measured as the average pretrial detention rate of the type of crime), an indicator for whether the individual was engaged in pretrial misconduct during a previous prosecution, an indicator for whether the individual has been convicted of a crime in the past, and type of crime fixed effects for the current prosecution (consisting of nine mutually exclusive categories). \textit{Attorney and judge controls} include judge leniency and public attorney quality and their squares, measured as residualized (against court-by-year fixed effects) leave out release rates, as in \cite{dgy}.}
}
\end{table}

\paragraph{Results} Table \ref{Tab_BT_main} presents the results. Each column represents a different regression and reports the estimated benchmark coefficient, $\widehat{\alpha}_D$, with its corresponding standard error. In the absence of OVB, $\widehat{\alpha}_D$ is a valid measure of discrimination, with negative values accounting for discrimination against immigrant defendants. Column 1 shows that the raw (uncontrolled) release rate for immigrants is 6.7 pp lower than the release rate for defendants that are Chilean nationals. The difference is quantitatively important: it represents 8\% of the unconditional average release rate or, put differently, almost 42\% of the unconditional average detention rate. 

If there are differences in misconduct potential between immigrant defendants and Chilean defendants, this result could be explained by OVB. Columns 2, 3, and 4 explore how the gap changes when adding the specific sets of controls previously discussed. When adding the individual controls (Column 2), the release rate gap increases to 10.3 pp. Under the assumption that criminal records and type of crime are positively correlated with misconduct potential, this suggests that the raw gap cannot be explained by differences in latent risk. In fact, the result conjectures that ignoring differences in misconduct potential may lead to an underestimation of the measure of discrimination. When adding the (residualized) controls for the assigned judge and attorney (Column 3), the release disparity remains unchanged. This is consistent with judges and attorneys being quasi-randomly assigned at the court-by-year level. When adding court-by-year fixed effects (Column 4), the release rate gap decreases to 3 pp. There are two potential explanations for this: The first explanation is that court-by-time fixed effects could mediate discrimination against immigrant defendants if they are systematically assigned to more severe courts (i.e., IVB). The second explanation is that the 3 pp decrease captures unobserved heterogeneity in misconduct potential that varies at the court-by-year level. Below we present suggestive evidence in favor of the first interpretation. Finally, when adding the three sets of controls (Column (5)), the gap in release rates between immigrant defendants and Chilean defendants is 8.7 pp. This represents 10.3\% of the unconditional average release rate or, put differently, 54.3\% of the unconditional average detention rate. We interpret these results as strong evidence of discrimination against defendants that are immigrants.

\subsection{Assessing Omitted Variable Bias}

The main threat to identification in benchmark regressions is OVB. If there are omitted variables that predict misconduct potential and that are correlated with immigration status, then the discrimination estimate will be biased. We conduct two sets of diagnostics to assess the extent to which OVB affects our main estimations. Both exercises suggest that unobservables that predict misconduct potential are unlikely to explain the results and, if anything, they are likely to exert downward bias on our discrimination estimates. The second diagnostic also provides elements for assessing the degree of IVB in our benchmark regressions.

\paragraph{Diagnostic 1: Bounds on selection on unobservables} One way of assessing the role of unobservables is to follow the extension proposed by \cite{oster2019unobservable} to the method proposed in \cite{altonji2005selection}. This approach interprets observables as unobservables that the econometrician happened to observe, establishing an implicit relationship between them. Under this interpretation, and assuming some degree of correlation between $X_i^o$ and $X_i^o$, the sensitivity of $\widehat{\alpha}_D$ to the inclusion of observed control variables is informative about potential OVB because the bias arising from omitting observables is related to the bias induced by the unobserved variables. This allows for the computation of bounds on $\widehat{\alpha}_D$ given different assumptions on the informativeness of observables about unobservables.

Based on \eqref{benchmark}, let $W_{1i} = X_i^{o\prime}\alpha_{Xo}$ and $W_{2i} = X_i^{u\prime}\alpha_{Xu}$, and define $\sigma_{jI} = \mathbb{C}(W_{ji},I_i)$ and $\sigma^2_j = \mathbb{V}(W_{ji})$. The intuition behind the method proposed in \cite{altonji2005selection} and extended by \cite{oster2019unobservable} is to assume that the correlation between the vectors $W_{ji}$ and $I_i$ is proportional---that is,
\begin{eqnarray}
\delta \frac{\sigma_{1I}}{\sigma_1^2} = \frac{\sigma_{2I}}{\sigma_2^2},
\end{eqnarray}
with $\delta$ being the coefficient of proportionality. This can be rationalized by assuming a correlation structure, in a projection sense, between the vectors $X_i^o$ and $X_i^u$. \cite{oster2019unobservable} assumes $\delta = 1$. That is to say, biases from omitting observables and unobservables are of similar magnitudes. Bounds, however, can be computed under different assumptions on $\delta$ to accommodate different a priori correlation structures.\footnote{Assuming $\delta = 1$ mimics a setting in which $X_i^o$ is a random subsample of $X_i$. If one assumes that the most important elements of $X_i$ are included in $X_i^o$, then $\delta$ is likely to be smaller than one.}

With an argument based on auxiliary regressions, \cite{oster2019unobservable} characterizes OVB as a function of $\delta$ that naturally suggests bounds on the estimated coefficient. \cite{oster2019unobservable} also notes that the degree to which the sensitivity analysis informs us about OVB depends on the variance of $\xi_i$ or on the contribution of $X_i$ to the variance of $Release_i$, conditional on $I_i$. Intuitively, coefficient stability may also arise from lack of explanatory power, so the implied OVB has to be scaled by the explanatory power of the observed and unobserved covariates. This implies that the econometrician also has to make an assumption about $R_{max}$---that is, the hypothetical $R^2$ that would be obtained after estimating \eqref{benchmark} by OLS including both $X_i^o$ and $X_i^u$. Bias is less likely to be problematic if coefficient stability after the inclusion of $X_i^o$ is accompanied by important increases in goodness of fit relative to $R_{max}$.

\begin{table}[t!]
\captionsetup{justification=centering,margin=0cm}
\RawCaption{\caption{\cite{oster2019unobservable} Test for Coefficient Stability: $\widehat{\alpha}_D$}\label{Tab_Oster}}

{}
{\small\begin{tabular}{rcccccc}\toprule
 & \multicolumn{6}{c}{$\delta$} \\
 & $-1$ & $-0.5$ &  $-0.25$ &  $0.25$ &  $0.5$ &  $1$  \\ \cline{2-7} 
 $R_{max}$ &  &  &  & &  &  \\ 
 $0.25$    &       -0.081   &        -0.084  &       -0.085    &        -0.088  &        -0.090 &        -0.093 \\ [2pt]
 $0.5$     &       -0.056   &        -0.071  &       -0.079    &        -0.095  &        -0.103 &        -0.121 \\ [2pt]
 $0.75$    &       -0.033   &        -0.059  &       -0.072    &        -0.102  &        -0.117 &        -0.151 \\ [2pt]
 $1$       &       -0.012   &        -0.047  &       -0.066    &        -0.109  &        -0.132 &        -0.185 \\ [4pt]
\bottomrule
\end{tabular}
\captionsetup{justification=justified,margin=0cm}
\floatfoot{\footnotesize\textbf{Notes:} This table presents bounds for the estimated coefficient $\widehat{\alpha}_D$ that correct for potential OVB by implementing the method developed by \cite{oster2019unobservable}, under different assumptions on $\delta$ and $R_{max}$. $\delta$ is the coefficient of proportionality that accounts for the correlation between observables and unobservables. $\delta = 0$ is not included because it rules out OVB by construction. $R_{max}$ is the $R^2$ of a regression that includes observables and unobservables.}
}
\end{table}

Table \ref{Tab_Oster} presents the estimated bounds for the coefficient $\widehat{\alpha}_D$ under different assumptions on $\delta$ and $R_{max}$. Each cell shows the estimated $\widehat{\alpha}_D$ parameter adjusted for OVB under the assumption that selection on observables informs us about selection on unobservables. Note that larger values for $\delta$ and $R_{max}$ imply more conservative bounds on $\widehat{\alpha}_D$ because both impose a stronger role for unobservables in the release equation. 

Imposing $\delta>0$, which is the standard assumption, suggests that discrimination is underestimated: bounds are larger (in absolute value) than the fully controlled benchmark regressions displayed in Column 5 of Table \ref{Tab_BT_main}. This follows from the fact that, on average, immigrant defendants have cleaner criminal records than Chilean defendants. It also relates to the discussion on Column 2 in Table \ref{Tab_BT_main}, which posited that, if anything, unobservables exert downward bias on the discrimination estimates. The results show that the only way in which OVB works against finding discrimination is by assuming that $X_i^o$ and $X_i^u$ are negatively correlated---that is, that defendants with observables that induce lower misconduct potential also have unobservables that work in the opposite direction. Although it is counterintuitive, this scenario can be explored by assuming negative values for $\delta$. Note that, even the in the most conservative (and presumably implausible) scenario in which the negative correlation is large ($\delta = -1$) and the explanatory power of omitted variables is sizable ($R_{max} = 1$), the point estimate still suggests (mild) discrimination against immigrant defendants. These results suggest that our main findings are unlikely to be driven by unobservables that explain misconduct potential and correlate with immigration status. By contrast, we provide suggestive evidence that, if anything, the results in Table \ref{Tab_BT_main} underestimate the level of discrimination against immigrant defendants in pretrial release decisions. 

\paragraph{Diagnostic 2: Using the outcome equation to test for omitted variable bias} We develop an alternative test to assess the role of unobservables in explaining the estimated release disparities. This approach is based on our theoretical framework and uses the estimation of the outcome equation (controlling for selection bias) to validate the benchmark equation. Intuitively, relevant omitted variables predict misconduct potential and correlate with immigration status. A regression of observed pretrial misconduct against immigration status controlling for our proxies of misconduct potential should therefore inform about the partial correlation between being an immigrant and pretrial misconduct. 

If we assume linear models for simplifying the exposition, then the release and outcome equations can be written as
\begin{eqnarray}
Release_i &=& \alpha_0 + \alpha_D I_i +X_i^{o\prime}\alpha_{Xo}  + X_i^{u\prime}\alpha_{Xu}  +  \xi_i,\label{R_eq_A} \\
PM_i &=& \beta_0 + \beta_D I_i + X_i^{o\prime}\beta_{Xo}  + X_i^{u\prime}\beta_{Xu}  +  \epsilon_i. \label{P_Meq_A}
\end{eqnarray}

In what follows, the estimated coefficients from regressions that include $(I_i,X_i^o,X_i^u)$ are denoted by $(\widehat{\alpha}_{D}^{ou},\widehat{\beta}_{D}^{ou})$, and the estimated coefficients from regressions that only include $(I_i,X_i^o)$ are denoted by $(\widehat{\alpha}_{D}^{o},\widehat{\beta}_{D}^{o})$. Concerns about OVB in observational benchmark regressions (possibly) imply that $\mathbb{E}\left[\widehat{\alpha}^{ou}_D|I_i,X_i\right]\neq \mathbb{E}\left[\widehat{\alpha}^{o}_D|I_i,X_i^o\right]$. We argue that failing to reject $\beta_{D} = 0$ in an outcome regression that omits $X_i^u$ implies that $\mathbb{E}\left[\widehat{\alpha}^{ou}_D|I_i,X_i\right]= \mathbb{E}\left[\widehat{\alpha}^{o}_D|I_i,X_i^o\right]$; thus, the outcome regression can be used as an indirect test for OVB in the release equation.

To see why, recall that, by assumption, $X_i = [X_i^o\quad X_i^u]$ is an accurate proxy of misconduct potential, implying that $\beta_D = 0$: immigration status cannot predict pretrial misconduct after controlling for misconduct potential. The hypothetical (unselected) regression of $PM_i$ versus $I_i$ and $X_i$ yields unbiased estimates. Specifically, $\mathbb{E}\left[\widehat{\beta}_D^{ou}|I_i,X_i\right] = \beta_D = 0$.\footnote{This requires us to assume that $\epsilon_i$ is uncorrelated with $I_i$, so $\mathbb{E}[\epsilon_i|I_i,X_i]=\mathbb{E}[\epsilon_i|X_i] = 0$. If we think of $\epsilon_i$ as the projection error, $Y_i^* = g(X_i) + \epsilon_i$, then the second equality is true by construction, and the first equality requires us to assume that the projection error does not vary with immigration status, which is a plausible assumption given that $X_i^u$ is unrestricted (eventually making $\epsilon_i$ negligible) and allowed to correlate with $I_i$.} This logic does not hold in a regression that omits $X_i^u$ because $\mathbb{E}\left[\widehat{\beta}_D^{o}|I_i,X_i^o\right] =\beta_D + bias$, where the bias depends on $\beta_{Xu}$ and the correlation between $I_i$ and $X_i^u$ through a standard OVB formula. Assuming $\beta_{Xu}\neq0$ (the contrary would imply that OVB is ruled-out by assumption), nonzero estimates of $\widehat{\beta}_D^o$ indicate that the correlation between $I_i$ and $X_i^u$ is nonzero. Intuitively, $I_i$ becomes a good predictor for pretrial misconduct only if it correlates with elements of $X_i$ that are not observed by the econometrician. 

The nonzero correlation between $I_i$ and $X_i^u$ is exactly what induces bias when estimating discrimination in the release equation without observing $X_i^u$. In particular, $\mathbb{E}\left[\widehat{\alpha}_D^{ou}|I_i,X_i\right] = \mathbf{D}$ because controlling by $X_i$ implies that $\widehat{\alpha}_D^{ou}$ identifies the direct effect of $I_i$, as well as the effects of other unobservables that are correlated with $I_i$ but that are not related with misconduct potential (contained in $\xi_i$) and, therefore, constitute valid sources of discrimination under a disparate impact perspective. By contrast, $\mathbb{E}\left[\widehat{\alpha}_D^{o}|I_i,X_I^o\right] = \textbf{D} + bias$, where the bias depends on $\alpha_{Xu}$ and, again, the correlation between $I_i$ and $X_i^u$. Note that the correlation that induces OVB in the release equation is the same that induces nonzero $\widehat{\beta}_D^o$ coefficients in the outcome equation.

Putting the two arguments together, we note that failing to reject $\beta_{D} = 0$ in a regression that omits $X_i^u$ (i.e., when the point estimate is $\hat\beta_{D}^{o}$) implies $\mathbf{D} = \mathbb{E}\left[\widehat{\alpha}^{ou}_D|I_i,X_i\right]= \mathbb{E}\left[\widehat{\alpha}^{o}_D|I_i,X_i^o\right]$. Put simply, not rejecting that immigration status does not predict pretrial misconduct on top of $X_i^o$ confirms that the vector $X_i^o$ successfully controls for group differences in misconduct potential and, therefore, the discrimination estimates from equation \eqref{R_eq_A} are not affected by standard OVB. One challenge in implementing this test is that the sample of released defendants is selected because pretrial misconduct if released is not observed for detained individuals. Accordingly, we estimate selection models that use the judge and attorney controls as the excluded variables in the outcome equation based on the assumption that judges' leniency and attorneys' quality affect selection into treatment but do not affect misconduct potential.\footnote{To avoid saturating the nonlinear first-stage model with the court-by-time fixed effects, we replace them with time-varying variables at the court level: number of judges, average pretrial release rate, and number of prosecutions (within a court in a given year). Estimating the fully controlled benchmark regression (Column 5 in Table \ref{Tab_BT_main}) and replacing the court-by-time fixed effects by these time-varying court-level controls yields a point estimate of -0.084, suggesting that the time-varying variables do a good job of approximating the court-by-time fixed effects.}\footnote{Although the test proposed here is different from the weighting correction made in \cite{arnold2020measuring}, it is worth noting that their proposal is also based on the intuition that the unselected outcome equation can be used to assess the degree of OVB in the selection equation.} Our main specification implements a standard two-step parametric Heckit selection correction \citep{heckman1974shadow}. We also provide results using the semiparametric correction proposed by \cite{newey2009two}.\footnote{The Heckit selection correction assumes that the error terms of the first and second steps are jointly normal. The semiparametric correction of \cite{newey2009two} 
uses series approximations to compute control function corrections. We implement the semiparametric correction following \cite{low2015disability} where the first step uses \cite{gallant1987semi} estimator to approximate the unknown density by third degree Hermite polynomial expansions and the second step controls for non-linear transformations of the density prediction. As in \cite{low2015disability}, we consider three models. Let $\hat{f}$ denote the predicted density. The control function used in Model I is $\hat{f}$ and its square, in Model II is $\Phi\left(\hat{\alpha}_0+\hat{\alpha}_1\hat{f}\right)$ and its square --where $\Phi$ is the normal cumulative distribution function and $\left(\hat{\alpha}_0,\hat{\alpha}_1\right)$ are the estimated coefficients of a Probit model of $Release$ on a constant and $\hat{f}$--, and in Model III is $\lambda\left(\hat{\alpha}_0+\hat{\alpha}_1\hat{f}\right)$ and its square --where $\lambda(x) = \phi(x)/\Phi(x)$ is the inverse Mills ratio and $\phi$ the normal density. }

\begin{table}[t!]
\captionsetup{justification=centering,margin=0cm}
\RawCaption{\caption{Parametric Selection Model for Assessing OVB: $\widehat{\beta}_D^o$}\label{Tab_OV_test}}
{}
{\small\begin{tabular}{lcccc}\toprule
 & (1) & (2) & (3) & (4) \\ \cline{2-5}
 &  &  &  &   \\ [-1pt]
 Immigrant                                   &       -0.078              &      -0.012  &       -0.090     &       -0.008        \\ [2pt]
                                             & (0.007)    & (0.006)    & (0.007)    & (0.006) \\ [5pt]
 Mean dep. variable                          &         0.29 &         0.29 &         0.29 &         0.29  \\ [5pt]
Individual controls                         & No & Yes   & No  & Yes \\ [3pt]
 Court-by-year controls               & No & No  & Yes   & Yes \\ [3pt]
  No. of Immigrants     &        4,900  &        4,900  &        4,900   &        4,900 \\ [2pt]
 No. of Chileans       &      580,406 &      580,406 &      580,406  &      580,406  \\ [2pt]
 F-test for excluded variables            &        969.6   &        969.6   &        969.6    &        969.6    \\ [2pt]
\bottomrule
\end{tabular}
\captionsetup{justification=justified,margin=0cm}
\floatfoot{\footnotesize\textbf{Notes:} This table presents the results from the two-step parametric sample selection model \citep{heckman1974shadow} with different sets of controls, where the dependent variable is pretrial misconduct. We report the point estimate for the immigrant indicator (i.e., the coefficient $\widehat{\beta}_D^{o}$) of equation \eqref{P_Meq_A} and its standard error. Both sets of controls (\textit{individual controls} and \textit{court-by-year controls}) are always included in the selection equation, but the columns vary in their inclusion in the outcome equation. Judge and attorney controls are defined as in Table \ref{Tab_BT_main} and are excluded from the outcome equation. The last row presents the joint F-test for judge and attorney (excluded) controls in the selection equation. \textit{Individual controls} are defined as in Table \ref{Tab_BT_main}. To avoid saturating the nonlinear first-stage with court-by-year fixed effects they are replaced in the regressions by court-by-year time varying covariates---namely, the average number of judges, the average pretrial release rate, and the number of prosecutions (within a court in a given year).}
}
\end{table}

Table \ref{Tab_OV_test} presents the results using the parametric selection correction. Table \ref{Tab_OV_NomPtest} of Appendix \ref{add} presents the results using the semiparametric selection correction. Each column represents a different regression and reports the estimated $\widehat{\beta}_D^{o}$ coefficient from equation \eqref{P_Meq_A} with its corresponding standard error. In all regressions, the selection equation includes the complete set of controls, but the included covariates in the outcome equation vary between columns. Note that the F-test for judge leniency and attorney quality (and their squares) suggests they effectively help with the selection correction. 

Column 1 in Table \ref{Tab_OV_test} shows that, after controlling for selection, pretrial misconduct rates are 7.8 pp smaller for immigrant defendants. This suggests that the benchmark regression with no controls (Column 1 in Table \ref{Tab_BT_main}) is affected by OVB. The implied OVB, however, suggests that abstracting from controls leads to an underestimation of the discrimination against immigrant defendants. Column 2 shows that including individual controls (criminal records and type of crime) reduces the pretrial misconduct rate differences to 1.2 pp. That is to say, conditioning on the individual controls reduces group variation in $Y_i^*$ by 85\%. This suggests that, although imperfect, criminal records and type of crime are good predictors of pretrial misconduct potential.  Adding time-varying court controls do little to correct for unobserved differences in misconduct potential (Column 3); however, adding both individual and time-varying court controls reduces the misconduct rate disparities to a nonsignificant gap of less than a percentage point. Taken together, we interpret this as support for our choice of $X_i^o$. The vector $X_i^o$ seems to do a good job in accounting for group differences in misconduct potential, which implies that the discrimination estimates of Column 5 of Table \ref{Tab_BT_main} are not driven by unobserved variation in latent risk. The model with the semiparametric selection correction proposed by \cite{newey2009two} yield the same conclusions.\footnote{The three models reproduce the pattern displayed in Table \ref{Tab_OV_test}. After accounting for selection, immigrants have lower pretrial misconduct rates. These gaps are substantially reduced after including individual controls in the outcome equation. In Model I, individual controls reduce the gap by two thirds, although the gap remains significant at 3 pp after including the complete set of controls. In Models II and III the individual controls completely close the gap in pretrial misconduct to a nonsignificant estimate of less than a percentage point, even in the specification that excludes time-varying court characteristics.}

A similar test can be implemented by comparing the role of unobservables in the selection and outcome equations using Kitagawa–Oaxaca–Blinder (KOB) decompositions \citep{kitagawa1955components,oaxaca1973male,blinder1973wage}. Intuitively, unobservables can affect the interpretation of the estimated benchmark coefficients if they correlate with immigration status and explain misconduct potential. If unobservables are important for predicting group-specific misconduct potential when making release decisions, they should also explain differences in observed pretrial misconduct rates.\footnote{One caveat of this analysis is that the sample used in the outcome equation is selected, presumably based on unobservables. To the extent that the selected sample has variation in unobservables, however, the exercise should still capture their effects. This would not be the case if, for example, the release decision is based on a binary unobservable such that all released defendants have the same value. In this case, the lack of variation in the selected sample would prevent the decomposition from capturing the effect of unobservables---that is, the KOB exercise could be noninformative under specific patterns of selection on unobservables.} Appendix \ref{KOB} shows KOB decompositions based on the vectors of individual covariates, using both raw data and residualized data against court-by-time fixed effects. The share of variation that is not explained by observables is large in the release equation but it is essentially zero in the outcome equation. This suggests that the immigrant indicator in the release equation is effectively capturing discrimination and not differences in unobservables.

\paragraph{On included variable bias} Our diagnostics do not explicitly deal with IVB concerns, but the second OVB diagnostic provides some insights to assess the pervasiveness of IVB. Individual controls are much more important than court-by-time variation in terms of accounting for group differences in pretrial misconduct rates. This indicates that the inclusion of court-by-time fixed effects in the benchmark regression may induce IVB in the sense that these particular fixed effects impact the estimated discrimination coefficient but do little to attenuate OVB. In other words, regressions that include court-by-year fixed effects could be over controlling and therefore inducing bias to our discrimination estimates.

IVB affects the interpretation of the sensitivity analysis. As shown in Table \ref{Tab_BT_main}, including court-by-year fixed effects significantly reduces the discrimination estimate. The results presented in Table \ref{Tab_OV_test} suggest that the sensitivity of $\widehat{\alpha}_D$ to the inclusion of court-by-year fixed effects is not explained by unobserved variation in misconduct potential but rather by an alternative mechanism that mediates discrimination against immigrant defendants. For example, the fact that discrimination decreases when controlling by court-by-time fixed effects may indicate that part of the discrimination against immigrant defendants is driven by their systematic allocation to courts that are more severe on average. Therefore, from a disparate impact perspective, the relevant discrimination estimate could be the one that only includes individual controls (Column 2 of Table \ref{Tab_BT_main}), which suggests larger levels of discrimination against immigrant defendants than the fully controlled specification. Because this conjectured IVB attenuates the discrimination estimate, we take a conservative position by presenting the fully controlled benchmark regression as our preferred specification. It is important to keep in mind, however, that the impact of IVB may mean that the actual discrimination against immigrant defendants is even larger.

\section{Dissecting the Discrimination Estimates}
\label{sec6}

Section \ref{sec5} documents a robust pretrial release rate disparity between immigrant defendants and Chilean defendants after controlling for variables that proxy for misconduct potential that we interpret as evidence of discrimination against immigrant defendants. This section dissects this reduced-form result to get a more comprehensive characterization of the discrimination patterns against immigrant defendants. All results presented in this section consider the full set of controls and should therefore be compared to Column 5 in Table \ref{benchmark}.

First, we explore some behavioral drivers of the discrimination estimate. We show that a sizable portion of the effect stems from an informational problem: immigrant criminal records are censored and this leads judges to punish immigrants with no previous prosecutions. Outcome test estimations show that the result cannot be rationalized by accurate statistical discrimination, suggesting that biased beliefs and taste-based discrimination play a role in the resolution of the informational problem. Second, we explore heterogeneities by type of crime and find that discrimination is especially large for drug-related offenses. This is consistent with a hostile discourse in the public sphere against immigrants that links immigration to drug trafficking. Finally, we explore the time trend of discrimination patterns considering the recent immigration wave discussed in Section \ref{sec3}. We find that discrimination has dramatically increased in recent years, and that this increase is not explained by the change in the composition of the immigrant population. 

\subsection{Behavioral Drivers of Discrimination}

The criminal records of immigrant defendants are censored (i.e., judges cannot observe the criminal histories of immigrant defendants in their origin countries). This means that judges may interpret the information available to them differently. In Bayesian jargon, the quality of the signal may differ between Chilean defendants and immigrant defendants. This is likely to be more detrimental to immigrant defendants with no criminal history in Chile because the main information contained in the criminal records is at the extensive margin. We explore whether this informational gap is in fact driving discrimination patterns against immigrant defendants.

\begin{table}[t!]
\captionsetup{justification=centering,margin=0cm}
\RawCaption{\caption{Behavioral Drivers of Discrimination and the Role of Information}\label{Tab_info_Hypothesis}}
{}
{\small\begin{tabular}{lccccc}\toprule
 & \multicolumn{3}{c}{Benchmark Test} &  \multicolumn{2}{c}{Outcome Test}   \\ \cline{2-4} \cline{5-6}
 & \multicolumn{2}{c}{Previous Prosecution} &  &  &  \\  
 & Yes & No &  & KPT & GV     \\ 
 & (1) & (2) & (3) & (4) & (5)  \\ \cline{2-6}
 &  &  &  & &  \\ [-1pt]
 Immigrant                                   &       -0.019   &        -0.132  &       -0.135    &       -0.050  & -0.074\\ [2pt]
                                             & (0.007) & (0.015) & (0.023)  & (0.007) & (0.016) \\ [3pt]
 Immigrant x experienced judge             &         &         &        0.049    &         &\\ [2pt]
                                             &         &         & (0.022)  &         &\\ [5pt]
 Mean dep. variable                          &        0.80 &         0.92    &        0.82 &     0.29     & 0.37\\ [6pt]
 No. of Immigrants  &        2,630 &        3,732 &        4,187 & 4,900          & 840\\ [2pt]
 No. of Chileans       &      472,636 &      220,370 &      357,520 &     580,406  &  57,690 \\ [2pt]
 R-squared             &         0.18 &         0.17 &         0.21 & & \\ [2pt]
\bottomrule
\end{tabular}
\captionsetup{justification=justified,margin=0cm}
\floatfoot{\footnotesize\textbf{Notes:} Columns 1 to 3 present the estimated benchmark coefficient, $\widehat{\alpha}_D$, of the extensions of equation \eqref{benchmark} and its corresponding standard error (clustered at the court-by-year level). Each column represents a different regression and includes all sets of controls analogous to Column 5 in Table \ref{Tab_BT_main} (see Table \ref{Tab_BT_main} notes for details). Columns 1 and 2 present $\widehat{\alpha}_D$ separately for defendants with and for defendants without previous prosecutions. Column 3 presents $\widehat{\alpha}_D$ for the period 2013--2017 and includes an interaction between $I_i$ and an indicator that takes the value one if the case was handled by an experienced judge (i.e., judges that handled 200 or more cases in the 2008--2012 period). Columns 4 and 5 present the outcome test estimations using \cite{kpt} (KPT) and \cite{pbot} (GV) approaches, respectively. KPT approach consists on an OLS regression of pretrial misconduct on the immigrant indicator for the complete sample of released defendants. GV performs a sample selection procedure that limits the extent of inframarginality bias in the KPT implementation. Details of the outcome test implementation of GV can be found in Appendix \ref{OT}.}}
\end{table}

To test the information hypothesis, we estimate the main benchmark regressions separately for defendants with and for defendants without previous prosecutions. These estimates include all sets of controls analogous to Column 5 in Table \ref{Tab_BT_main}. Results are displayed in Columns 1 and 2 in Table \ref{Tab_info_Hypothesis}. Immigrant defendants with previous prosecutions are 1.9 pp less likely to be released compared to Chilean defendants with previous prosecutions; compared to a difference of 13.2 pp for defendants without previous prosecutions. To assess the magnitude of the difference between the estimations, note that the (unconditional) pretrial release rate is 12 pp larger for defendants with clean criminal records, which is roughly similar in magnitude to the difference in the discrimination estimate between samples. This implies that not having previous prosecutions is, essentially, irrelevant for immigrant defendants: they are not ``rewarded'' for having clean criminal records. This result is consistent with judges systematically imputing nonzero criminal histories for all immigrant defendants regardless of their criminal record in Chile. 

To further explore this intuition, we test whether judge experience matters for the aggregate discrimination patterns. If judges face an informational problem whose imperfect solution leads to discriminatory practices, they should gradually implement nondiscriminatory solutions as they gain experience. Consequently, we should observe that more experienced judges exhibit lower levels of discrimination. To test this, we define experienced judges as those who handled more than 200 cases in the period 2008--2012 and estimate the benchmark regressions for the period 2013-2017, including the experience indicator and its interaction with the immigrant indicator. Column 3 shows that release disparities are around 35\% smaller for experienced judges.

As discussed in Section \ref{sec2}, aggregate patterns of discrimination represent a possibly complex combination of different behavioral sources---these being, statistical discrimination, biased beliefs, and taste-based discrimination, which is not possible to identify separately without additional structural assumptions \citep{arnold2020measuring,bohren2020inaccurate,hull2021}. The fact that discrimination in pretrial detention decisions stems from an informational problem does not necessarily mean that the aggregate estimate can be rationalized by accurate statistical discrimination. On the one hand, judges can circumvent the informational problem based on stereotypes. This may, for example, lead judges to ``overcorrect'' for censored criminal records for immigrants. On the other hand, the informational problem may provide judges with the room to exercise taste-based discrimination. The fact that experienced judges still display significant discrimination against immigrants suggests that these drivers are likely to play a role in the analysis.

To get additional insights, we implement the outcome test \citep{b1,b2} using the observational approaches of \cite{kpt} and \cite{pbot}.\footnote{The small share of the immigrant population prevents us from implementing the instrument-based approach proposed by \cite{ady}.} The outcome test is a diagnostic for differences in effective selection thresholds between groups. Using the notation set out in Section \ref{sec2} (equation \eqref{release2}), the test explores whether $\mathbb{E}[t_{j(i)}(1,Z_i) - b_{j(i)}(1,Z_i)]$ is different from $\mathbb{E}[t_{j(i)}(0,Z_i) - b_{j(i)}(0,Z_i)]$, where the expectation is taken across $Z_i$ and $j(i)$. Differences in effective thresholds can be identified by differences in observed misconduct among marginally released defendants \citep{ady,hull2021}. \cite{kpt} propose using the average behavior of released defendants. \cite{pbot} refine \cite{kpt} approach by performing a sample selection procedure that limits the extent of inframarginality bias.\footnote{When risk distributions vary by group, the behavior of the average released defendant is potentially different from the behavior of the marginally released defendants. This is called inframarginality bias. \cite{pbot} deal with this problem in two stages. First, it uses observational variation to identify samples of marginal defendants. The identification argument works under sufficient conditions which are supported by suggestive evidence presented in Appendix \ref{OT}. Second, it performs differences in means of pretrial misconduct rates between immigrant defendants and Chilean defendants using the samples of marginal defendants. The sample of marginal defendants is defined as the 10\% of released defendants with larger conditional probabilities of being marginal. That is why the number of observations is smaller when using this approach. Additional details about the implementation of the outcome test proposed by \cite{pbot} are presented in Appendix \ref{OT}.} 

Intuitively, discrimination in pretrial detention decisions may originate from judges setting stricter release thresholds for immigrant defendants, in a disparate impact sense, that imply lower release rates at equivalent true pretrial misconduct probabilities. These differences in the treatment of immigrant defendants could be driven by stereotypes or animus or both. This source of discrimination does not incorporate sources of release disparities driven by differences in group-specific misconduct potential. Differences in true pretrial misconduct potential affect how often a given defendant crosses the release threshold, but the threshold itself is not informative about them. Then, rejecting the outcome test implies that the observed behavior of marginally released defendants is inconsistent with a selection process that lacks stereotypes and animus \citep{hull2021}. 

Columns 4 and 5 in Table \ref{benchmark} show the results of the outcome test using \cite{kpt} and \cite{pbot} approaches, respectively. Estimates indicate that immigrant defendants face effective release thresholds that are, on average, between 5 and 7.2 pp stricter. That is to say, immigrants are required to have, on average, smaller pretrial misconduct probabilities to qualify for pretrial release relative to Chilean defendants. Although the magnitudes between benchmark and outcome regressions are not directly comparable \citep{EL}, this result rejects accurate statistical discrimination as the only driver of the discrimination estimates, implying that a combination of stereotypes and animus plays an important role in explaining aggregate patterns of discrimination against immigrants.\footnote{The average pretrial misconduct rate is larger in Column 5 (37\%) relative to Column 4 (29\%). This suggests that \cite{pbot} implementation of the outcome test successfully attenuates inframarginality concerns by focusing on a sample of released defendants that are, on average, riskier and, therefore, presumably more likely to be at the margin of pretrial detention.}

\subsection{Heterogeneity by Type of Crime}

If statistical structures, stereotypes, or judges preferences vary with type of crime, then discrimination patterns should do as well. Exploring this particular pattern of heterogeneity is instructive because a hostile facet of the public discourse regarding immigrants links immigration with drug trafficking and violent crimes, so discrimination should be particularly salient in pretrial detention decisions concerning these crimes.

To investigate this, we split prosecutions between the following nine mutually exclusive categories: homicides, sexual offenses, thefts and robberies, other property crimes, drug offenses, white collar crimes and tax crimes, crimes against public trust, crimes against the freedom and privacy of people, and other crimes.\footnote{For examples of the crimes included in each category, see Appendix \ref{data_app}.} Using these categories, we estimate the following regression:
\begin{eqnarray}
Release_i &=& \alpha_0 + \sum_{k=1}^9\alpha_D^k I_i\cdot C_i^k + X_i^{o\prime}\alpha_{Xo} + \varepsilon_i,\label{benchmark_drug}
\end{eqnarray}
where $C_i^k$ is an indicator that takes the value one if the imputed crime is in category $k$. All other variables are defined as in equation \eqref{benchmark}; therefore, the noninteracted crime indicators are included in $X_i^o$. In this specification, $\widehat{\alpha}_D^k$ identifies the controlled pretrial release disparities between Chilean defendants and immigrant defendants (our measure of discrimination) in crime category $k$.

\begin{table}[t!]
\captionsetup{justification=centering,margin=0cm}
\RawCaption{\caption{Discrimination by Type of Crime: $\widehat{\alpha}_D^k$}\label{Tab_BT_Crime_types}}
{}
{\small\begin{tabular}{lcc}\toprule
 & Coeff. & SE \\ \cline{2-3} 
 &  &   \\ [-1pt]
 Homicide                          &       -0.030  &  (0.067)  \\ [2pt]
 
 Sexual offense                               &        0.001  &  (0.033)  \\ [2pt]
 
  Theft or robbery                          &       -0.038  &  (0.018)  \\ [2pt]
 
  Other property crime                    &       -0.052  &  (0.014)  \\ [2pt]
 
 Drug offense                               &       -0.351  &  (0.034)  \\ [2pt]

White collar or tax crime                     &       -0.020  &  (0.019)  \\ [2pt]

 Crime against public trust                 &        0.021  &  (0.015)  \\ [2pt]
 Crime against people's freedom and privacy                      &       -0.007  &  (0.006)  \\ [2pt]
 
  Other crimes                                &        0.019  &  (0.021)  \\ [4pt]
 Mean dep. variable &         0.84 &  \\ [6pt]
 No. of Immigrants  &        6,362 &   \\ [2pt]
 No. of Chileans    &      693,006 &  \\ [2pt]
 R-squared          &        0.192 &  \\ [2pt]
\bottomrule
\end{tabular}
\captionsetup{justification=justified,margin=0cm}
\floatfoot{\footnotesize\textbf{Notes:} This table presents the estimated benchmark coefficients by type of crime, $\widehat{\alpha}_D^k$, of equation \eqref{benchmark_drug} and its corresponding standard error in parentheses (clustered at the court-by-year level). Each column represents a different regression and includes all sets of controls analogous to Column 5 in Table \ref{Tab_BT_main} (see Table \ref{Tab_BT_main} notes for details).}
}
\end{table}

Table \ref{Tab_BT_Crime_types} shows the results. Discrimination against immigrants is especially large for drug offenses---around four times larger than the baseline benchmark estimate. Significant effects are also found for thefts and robberies and property crimes. Other violent crimes like homicides and sexual offenses show no significant differences; however, as shown in Table \ref{Tab_Crime_type} of Appendix \ref{add}, the number of immigrants imputed for sexual offences and homicides in the period considered is too low, therefore potentially being underpowered to precisely estimate disparities in those particular categories. Crime categories that are associated with less hostile discourses in the public sphere (such as white collar crimes and tax crimes or crimes against privacy, and for which the number of prosecuted immigrants is in fact larger) also show no significant release disparities. Overall, these results suggest that discrimination patterns are mainly driven by drug offenses, property crimes, and thefts and robberies. This is consistent with idea that public discourses regarding immigrants play a role in the formation of stereotypes and hostile preferences.

\subsection{Time Trend and the Immigration Wave} 

Recall in Section \ref{sec3} it was premised that immigration flows have substantially increased in recent years, generating a hostile discourse against immigrants in the public sphere. It was also argued that the composition of the immigrant population by country of origin has changed considerably. We now examine whether these changes have been accompanied by a change in the aggregate discrimination patterns across time.

\begin{table}[t!]
\captionsetup{justification=centering,margin=0cm}
\RawCaption{\caption{Discrimination in Different Periods and Group Heterogeneity: $\widehat{\alpha}_D$}\label{BT_Time}}
{}
{\small\begin{tabular}{lcccc}\toprule
 & \multicolumn{2}{c}{2008-2012} & \multicolumn{2}{c}{2013-2017}   \\
 & (1) & (2) & (3) & (4)   \\ \cline{2-5}
 &  &  &  &   \\ [-1pt]
 Immigrant                                   &       -0.066   &             &       -0.101    &           \\ [2pt]
                                             & (0.016) &             & (0.015)  &           \\ [3pt]
 Immigrant (old group)                       &         &        -0.064     &          &        -0.099   \\ [2pt]
                                             &         & (0.017)    &          &  (0.015) \\ [3pt]
 Immigrant (new group)                       &         &             &          &        -0.108   \\ [2pt]
                                             &         &             &          & (0.022)  \\ [5pt]
 Mean dep. variable                          &        0.85 &         0.85    &        0.82 &         0.82  \\ [6pt]
 No. of Immigrants (old group)     &        1,956 &        1,956 &        2,873 &        2,873     \\ [2pt]
 No. of Immigrants (new group)     &          219 &            0 &        1,314 &        1,314     \\ [2pt]
 No. of Chileans       &      335,486 &      335,486 &      357,520 &      357,520   \\ [2pt]
 R-squared             &        0.172 &        0.172 &        0.207 &        0.207  \\ [2pt]
\bottomrule
\end{tabular}
\captionsetup{justification=justified,margin=0cm}
\floatfoot{\footnotesize\textbf{Notes:} This table presents the estimated benchmark coefficient, $\widehat{\alpha}_D$, of extensions for equation \eqref{benchmark} and its corresponding standard error (clustered at the court-by-year level). Each column represents a different regression and includes all sets of controls analogous to Column 5 in Table \ref{Tab_BT_main} (see Table \ref{Tab_BT_main}  notes for details). Columns 1 and 2 use data from the period 2008--2012, and Columns 3 and 4 use data from the period 2013--2017. In columns 2 and 4 the immigrant indicator is defined separately depending on the nationality of the immigrant. \textit{New group} includes immigrants from Colombia, Haiti, and Venezuela. \textit{Old group} considers immigrants from all the other countries.}
}
\end{table}

Following the changes in immigration flows described in Section \ref{sec3}, Table \ref{BT_Time} shows the benchmark estimates separately for the periods 2008--2012 and 2013--2017. The fully controlled discrimination estimate increased from 6.6 to 10.1 pp, a 53\% increase. A potential explanation for this increase is the increase in the hostility of the public discourse that may have affected stereotypes and preferences. Alternatively, this result could be driven by a composition effect. Although Peru was the principal source of immigration flows to Chile for many years, beginning in 2013 the relative share of immigrants from Colombia, Haiti, and Venezuela increased considerably. If there is heterogeneity in discrimination patterns---that is, discrimination is more intense against the ``new'' group of immigrants (for example, because of more salient racial differences)---then the increase of discrimination between the two periods could be explained by a composition effect and not by an increase in the average intensity of discrimination for a given nationality.\footnote{Potential differences in risk between the two groups are not a concern for the dynamic results because benchmark regressions include the full set of controls.} Column 4 presents the discrimination estimate for the 2013--2017 period separately for the ``new'' and the ``old'' groups of immigrants.\footnote{This exercise cannot be computed for the 2008--2012 period because the ``new'' group, by definition, is essentially nonexistent before 2013, as the number of observations suggest.} We cannot reject that the discrimination estimates for both groups are equal, which suggests that the increase in discrimination is unlikely to be driven by the change in the composition of the immigrant population.

\section{Conclusion}
\label{sec7}

This paper leverages rich administrative data to test for discrimination in pretrial detention decisions against defendants who are immigrants in Chile. We find that immigrant defendants are released pretrial at substantially lower rates relative to Chilean defendants with similar criminal records, case characteristics, and court-by-time fixed effects. Several diagnostics suggest that these estimated disparities are not driven by unobserved differences in pretrial misconduct potential between Chilean defendants and immigrant defendants. We therefore interpret the results as robust evidence of discrimination against immigrant defendants in pretrial detention decisions.

We show that release disparities are an order of magnitude larger for defendants with no previous criminal prosecutions in Chile. We argue this implies that discrimination is originated in an informational problem driven by censored criminal records (or, put differently, weaker signal quality). However, we show that accurate statistical discrimination cannot rationalize the release disparities, suggesting that  biased beliefs and taste-based discrimination play an important role. Consistent with the hostile discourses against immigrants in the public sphere, we also find that discrimination in pretrial detention decisions is particularly severe for drug offenses, and that the intensity of discrimination has increased in recent years during which the magnitude and the composition of the immigrant population has changed considerably.

Overall, our findings suggest that discrimination against immigrant defendants in pretrial detention decisions is substantial in the Chilean context and highlight the importance of implementing effective anti-discrimination policies to better assimilate the integration of immigrants in destination countries. Several countries are enacting anti-discrimination policies (or strengthening existing ones) to foster the integration of immigrant populations \citep{OECD}. This policy challenge appears to be particularly urgent in countries where immigration is a relatively recent phenomenon and institutions are not adequately equipped to deal with unexpected shocks in this regard.  

\bibliographystyle{apalike}
\bibliography{referencias}

\begin{thebibliography}{}

\bibitem[Abrams et~al., 2012]{abrams2012judges}
Abrams, D.~S., Bertrand, M., and Mullainathan, S. (2012).
\newblock Do judges vary in their treatment of race?
\newblock {\em The Journal of Legal Studies}, 41(2):347--383.

\bibitem[Aigner and Cain, 1977]{aigner1977statistical}
Aigner, D.~J. and Cain, G.~G. (1977).
\newblock Statistical theories of discrimination in labor markets.
\newblock {\em Industrial and Labor Relations Review}, 30(2):175--187.

\bibitem[Ajzenman et~al., 2021]{ajzenman2021immigration}
Ajzenman, N., Dominguez, P., and Undurraga, R. (2021).
\newblock Immigration, crime, and crime (mis) perceptions.

\bibitem[Alesina et~al., 2021]{alesina2021immigration}
Alesina, A., Harnoss, J., and Rapoport, H. (2021).
\newblock {Immigration and the future of the welfare state in Europe}.
\newblock {\em The Annals of the American Academy of Political and Social
  Science}, 697(1):120--147.

\bibitem[Alesina et~al., 2022]{alesinastant}
Alesina, A., Miano, A., and Stantcheva, S. (2022).
\newblock Immigration and redistribution.
\newblock {\em Review of Economic Studies}.

\bibitem[Altonji et~al., 2005]{altonji2005selection}
Altonji, J.~G., Elder, T.~E., and Taber, C.~R. (2005).
\newblock Selection on observed and unobserved variables: Assessing the
  effectiveness of catholic schools.
\newblock {\em Journal of Political Economy}, 113(1):151--184.

\bibitem[Arnold et~al., 2021]{arnold2020measuring}
Arnold, D., Dobbie, W., and Hull, P. (2021).
\newblock {Measuring racial discrimination in bail decisions}.
\newblock {\em Working Paper}.

\bibitem[Arnold et~al., 2018]{ady}
Arnold, D., Dobbie, W., and Yang, C. (2018).
\newblock {Racial bias in bail decisions}.
\newblock {\em Quarterly Journal of Economics}, 133(4):1885--1932.

\bibitem[{\AA}slund et~al., 2014]{aaslund2014seeking}
{\AA}slund, O., Hensvik, L., and Skans, O.~N. (2014).
\newblock Seeking similarity: How immigrants and natives manage in the labor
  market.
\newblock {\em Journal of Labor Economics}, 32(3):405--441.

\bibitem[Becker, 1957]{b1}
Becker, G. (1957).
\newblock {\em The Economics of Discrimination}.
\newblock University of Chicago Press.

\bibitem[Becker, 1993]{b2}
Becker, G. (1993).
\newblock {Nobel Lecture: The economic way of looking at behavior}.
\newblock {\em Journal of Political Economy}, 101:385--409.

\bibitem[Blinder, 1973]{blinder1973wage}
Blinder, A.~S. (1973).
\newblock {Wage discrimination: Reduced form and structural estimates}.
\newblock {\em Journal of Human resources}, 8(4):436--455.

\bibitem[Bohren et~al., 2021]{bohren2020inaccurate}
Bohren, J.~A., Haggag, K., Imas, A., and Pope, D.~G. (2021).
\newblock {Inaccurate statistical discrimination: An identification problem}.
\newblock {\em Working Paper}.

\bibitem[Bordalo et~al., 2016]{bordalo2016stereotypes}
Bordalo, P., Coffman, K., Gennaioli, N., and Shleifer, A. (2016).
\newblock Stereotypes.
\newblock {\em The Quarterly Journal of Economics}, 131(4):1753--1794.

\bibitem[Brell et~al., 2020]{brell2020labor}
Brell, C., Dustmann, C., and Preston, I. (2020).
\newblock The labor market integration of refugee migrants in high-income
  countries.
\newblock {\em Journal of Economic Perspectives}, 34(1):94--121.

\bibitem[Bursztyn et~al., 2021a]{bursztyn2021immigrant}
Bursztyn, L., Chaney, T., Hassan, T.~A., and Rao, A. (2021a).
\newblock The immigrant next door: Exposure, prejudice, and altruism.
\newblock {\em Working Paper}.

\bibitem[Bursztyn et~al., 2021b]{bursztyn2021disguising}
Bursztyn, L., Haaland, I.~K., Rao, A., and Roth, C.~P. (2021b).
\newblock Disguising prejudice: Popular rationales as excuses for intolerant
  expression.
\newblock {\em Working Paper}.

\bibitem[Cort\'{e}s et~al., 2019]{Cortes_et_al2019}
Cort\'{e}s, T., Grau, N., and Rivera, J. (2019).
\newblock {Juvenile incarceration and adult recidivism}.
\newblock {\em Working Paper}.

\bibitem[Dobbie et~al., 2018]{dgy}
Dobbie, W., Goldin, J., and Yang, C.~S. (2018).
\newblock The effects of pretrial detention on conviction, future crime, and
  employment: Evidence from randomly assigned judges.
\newblock {\em American Economic Review}, 108(2):201--240.

\bibitem[Dobbie and Yang, 2021a]{dobbie2021economic}
Dobbie, W. and Yang, C. (2021a).
\newblock The economic costs of pretrial detention.
\newblock {\em Brookings Papers on Economic Activity}.

\bibitem[Dobbie and Yang, 2021b]{dobbie2021jep}
Dobbie, W. and Yang, C. (2021b).
\newblock {The US pretrial system: Balancing individual rights and public
  interests}.
\newblock {\em Journal of Economic Perspectives}, 35(4):49--70.

\bibitem[Dom\'inguez et~al., 2022]{EL}
Dom\'inguez, P., Grau, N., and Vergara, D. (2022).
\newblock {Combining discrimination diagnostics to identify sources of
  statistical discrimination}.
\newblock {\em Economics Letters}, 212, 110294.

\bibitem[Dustmann et~al., 2017]{dustmann2017economics}
Dustmann, C., Fasani, F., Frattini, T., Minale, L., and Sch{\"o}nberg, U.
  (2017).
\newblock On the economics and politics of refugee migration.
\newblock {\em Economic policy}, 32(91):497--550.

\bibitem[Dustmann and Preston, 2007]{dustmann2007racial}
Dustmann, C. and Preston, I. (2007).
\newblock Racial and economic factors in attitudes to immigration.
\newblock {\em The BE Journal of Economic Analysis \& Policy}, 7(1).

\bibitem[Egger et~al., 2022]{dennis}
Egger, D., Auer, D., and Kunz, J. (2022).
\newblock Effects of migrant networks on labor market integration, local firms
  and employees.
\newblock {\em Working Paper}.

\bibitem[Fasani et~al., 2019]{fasani2019does}
Fasani, F., Mastrobuoni, G., Owens, E.~G., and Pinotti, P. (2019).
\newblock {\em Does Immigration Increase Crime?}
\newblock Cambridge University Press.

\bibitem[Gallant and Nychka, 1987]{gallant1987semi}
Gallant, A.~R. and Nychka, D.~W. (1987).
\newblock Semi-nonparametric maximum likelihood estimation.
\newblock {\em Econometrica}, 55(2):363--390.

\bibitem[Gelman et~al., 2007]{gelman2007analysis}
Gelman, A., Fagan, J., and Kiss, A. (2007).
\newblock {An analysis of the New York City police department's
  “stop-and-frisk” policy in the context of claims of racial bias}.
\newblock {\em Journal of the American Statistical Sssociation},
  102(479):813--823.

\bibitem[Grau et~al., 2021]{grau2019effect}
Grau, N., Marivil, G., and Rivera, J. (2021).
\newblock {The effect of pretrial detention on labor market outcomes}.
\newblock {\em Journal of Quantitative Criminology}.

\bibitem[Grau and Vergara, 2021]{pbot}
Grau, N. and Vergara, D. (2021).
\newblock An observational implementation of the outcome test with an
  application to ethnic prejudice in pretrial detentions.
\newblock {\em Working Paper}.

\bibitem[Grigorieff et~al., 2020]{grigorieff2020does}
Grigorieff, A., Roth, C., and Ubfal, D. (2020).
\newblock Does information change attitudes toward immigrants?
\newblock {\em Demography}, 57(3):1117--1143.

\bibitem[Hangartner et~al., 2019]{hangartner2019does}
Hangartner, D., Dinas, E., Marbach, M., Matakos, K., and Xefteris, D. (2019).
\newblock Does exposure to the refugee crisis make natives more hostile?
\newblock {\em American Political Science Review}, 113(2):442--455.

\bibitem[Heath et~al., 2013]{heath2013discrimination}
Heath, A., Liebig, T., and Simon, P. (2013).
\newblock {Discrimination against immigrants - Measurement, incidence and
  policy instruments}.
\newblock {\em International Migration Outlook - OECD}, pages 191--230.

\bibitem[Heaton et~al., 2017]{heaton2017downstream}
Heaton, P., Mayson, S., and Stevenson, M. (2017).
\newblock The downstream consequences of misdemeanor pretrial detention.
\newblock {\em Stanford Law Review}, 69:711.

\bibitem[Heckman, 1974]{heckman1974shadow}
Heckman, J. (1974).
\newblock Shadow prices, market wages, and labor supply.
\newblock {\em Econometrica}, 42(4):679--694.

\bibitem[Hopkins et~al., 2019]{hopkins2019muted}
Hopkins, D.~J., Sides, J., and Citrin, J. (2019).
\newblock The muted consequences of correct information about immigration.
\newblock {\em The Journal of Politics}, 81(1):315--320.

\bibitem[Hull, 2021]{hull2021}
Hull, P. (2021).
\newblock What marginal outcome tests can tell us about racially biased
  decision-making.
\newblock {\em Working Paper}.

\bibitem[J{\o}rgensen and Osmundsen, 2020]{jorgensen2020correcting}
J{\o}rgensen, F.~J. and Osmundsen, M. (2020).
\newblock Correcting citizens’ misperceptions about non-western immigrants:
  Corrective information, interpretations, and policy opinions.
\newblock {\em Journal of Experimental Political Science}, pages 1--10.

\bibitem[Kitagawa, 1955]{kitagawa1955components}
Kitagawa, E.~M. (1955).
\newblock Components of a difference between two rates.
\newblock {\em Journal of the American Statistical Association},
  50(272):1168--1194.

\bibitem[Knowles et~al., 2001]{kpt}
Knowles, J., Persico, N., and Todd, P. (2001).
\newblock {Racial bias in motor vehicle searches: Theory and evidence}.
\newblock {\em Journal of Political Economy}, 109(1):203--229.

\bibitem[Leslie and Pope, 2017]{leslie2017unintended}
Leslie, E. and Pope, N.~G. (2017).
\newblock {The unintended impact of pretrial detention on case outcomes:
  Evidence from New York City arraignments}.
\newblock {\em The Journal of Law and Economics}, 60(3):529--557.

\bibitem[Low and Pistaferri, 2015]{low2015disability}
Low, H. and Pistaferri, L. (2015).
\newblock Disability insurance and the dynamics of the incentive insurance
  trade-off.
\newblock {\em American Economic Review}, 105(10):2986--3029.

\bibitem[Newey, 2009]{newey2009two}
Newey, W.~K. (2009).
\newblock Two-step series estimation of sample selection models.
\newblock {\em The Econometrics Journal}, 12:S217--S229.

\bibitem[Oaxaca, 1973]{oaxaca1973male}
Oaxaca, R. (1973).
\newblock Male-female wage differentials in urban labor markets.
\newblock {\em International Economic Review}, 14(3):693--709.

\bibitem[OECD, 2019]{OECD}
OECD (2019).
\newblock International migration outlook.
\newblock {\em Organisation for Economic Co-operation and Development}.

\bibitem[Oster, 2019]{oster2019unobservable}
Oster, E. (2019).
\newblock Unobservable selection and coefficient stability: Theory and
  evidence.
\newblock {\em Journal of Business \& Economic Statistics}, 37(2):187--204.

\bibitem[Rehavi and Starr, 2014]{rehavi2014racial}
Rehavi, M.~M. and Starr, S.~B. (2014).
\newblock Racial disparity in federal criminal sentences.
\newblock {\em Journal of Political Economy}, 122(6):1320--1354.

\bibitem[Rose, 2022]{rose2022constructivist}
Rose, E.~K. (2022).
\newblock A constructivist perspective on empirical discrimination research.
\newblock {\em Working Paper}.

\bibitem[Stevenson and Mayson, 2021]{stevenson2021pretrial}
Stevenson, M.~T. and Mayson, S.~G. (2021).
\newblock Pretrial detention and the value of liberty.
\newblock {\em Working Paper}.

\bibitem[UN, 2013]{un2}
UN (2013).
\newblock International migration policies: Government views and priorities.
\newblock {\em United Nations - Population Division}.

\bibitem[UN, 2019]{un}
UN (2019).
\newblock International migration 2019 report.
\newblock {\em United Nations - Department of Economic and Social Affairs}.

\bibitem[Yang and Dobbie, 2020]{yang2020equal}
Yang, C. and Dobbie, W. (2020).
\newblock Equal protection under algorithms: A new statistical and legal
  framework.
\newblock {\em Michigan Law Review}.

\end{thebibliography}

\newpage

\appendix

\setcounter{footnote}{0} 
\begin{center}
\LARGE{Discrimination Against Immigrants in the Criminal\\ Justice System: Evidence from Pretrial Detentions}\\
\vspace{.5cm}
\Large{\textbf{Online Appendix}}\\
\Large{Patricio Dom\'inguez, Nicol\'as Grau, Dami\'an Vergara}
\end{center}
\thispagestyle{empty}

\vspace{4cm}
\startcontents[sections]
\printcontents[sections]{l}{1}{\setcounter{tocdepth}{2}}

\newpage
\pagenumbering{roman} 

\newpage
\section{Additional Figures and Tables}
\label{add}

\setcounter{equation}{0} \renewcommand{\theequation}{A.\Roman{equation}}

\setcounter{table}{0} \renewcommand{\thetable}{A.\Roman{table}}

\setcounter{figure}{0} \renewcommand{\thefigure}{A.\Roman{figure}}

\begin{table}[h]
\RawCaption{\caption{Semiparametric Selection Model for Assessing OVB: $\widehat{\beta}_D^o$}\label{Tab_OV_NomPtest}}
{\small\begin{tabular}{lcccc}\toprule
 & (1) & (2) & (3) & (4) \\ \cline{2-5}
 &  &  &  &   \\ [-1pt]
\textbf{Model I}: \ \ Immigrant             &       -0.091       &       -0.102       &       -0.029       &       -0.030        \\ [2pt]
                                            &       (0.0064)     &       (0.0064)     &       (0.0061)     &      (0.0061) \\ [5pt]
\textbf{Model II}: \ Immigrant              &       -0.081       &       -0.093       &       -0.006       &       -0.006        \\ [2pt]
                                            &       (0.0064)     &       (0.0064)     &       (0.0061)     &      (0.0061) \\ [5pt]
\textbf{Model III}: Immigrant               &       -0.080       &       -0.092       &       -0.005       &       -0.005        \\ [2pt]
                                            &       (0.0064)     &       (0.0063)     &       (0.0061)     &        (0.0061) \\ [5pt]
 Mean dep. variable                          &         0.29 &         0.29 &         0.29 &         0.29  \\ [5pt]
 Court-by-year characteristics               & No & Yes  & No   & Yes \\ [3pt]
 Individual controls                         & No & No   & Yes  & Yes \\ [3pt]
 No. of Immigrants     &        4,900  &        4,900  &        4,900   &        4,900 \\ [2pt]
 No. of Chileans       &      580,406 &      580,406 &      580,406  &      580,406  \\ [2pt]
\bottomrule
\end{tabular}
\captionsetup{justification=justified,margin=0cm}
\floatfoot{\scriptsize\textbf{Notes:} This table presents the results of the OVB test proposed in Section \ref{sec5} using the semiparametric correction of \cite{newey2009two} that
uses series approximations to compute control function corrections. We implement the semiparametric correction following \cite{low2015disability} where the first step uses \cite{gallant1987semi} estimator to approximate the unknown density by third degree Hermite polynomial expansions and the second step controls for non-linear transformations of the density prediction. As in \cite{low2015disability}, we consider three models. Let $\hat{f}$ denote the predicted density. The control function used in Model I is $\hat{f}$ and its square, in Model II is $\Phi\left(\hat{\alpha}_0+\hat{\alpha}_1\hat{f}\right)$ and its square --where $\Phi$ is the normal cumulative distribution function and $\left(\hat{\alpha}_0,\hat{\alpha}_1\right)$ are the estimated coefficients of a Probit model of $Release$ on a constant and $\hat{f}$--, and in Model III is $\lambda\left(\hat{\alpha}_0+\hat{\alpha}_1\hat{f}\right)$ and its square --where $\lambda(x) = \phi(x)/\Phi(x)$ is the inverse Mills ratio and $\phi$ the normal density. We report the point estimate for the immigrant indicator (i.e., the coefficient $\widehat{\beta}_D^{o}$) of equation \eqref{P_Meq_A} and its standard error. Standard errors are computed using bootstrap with 500 repetitions to account for the fact that the density is estimated in the first stage. Both sets of controls (\textit{individual controls} and \textit{court-by-year controls}) are always included in the selection equation, but the columns vary in their inclusion in the outcome equation. Judge and attorney controls are defined as in Table \ref{Tab_BT_main} and are excluded from the outcome equation. \textit{Individual controls} are defined as in Table \ref{Tab_BT_main}. To avoid saturating the nonlinear first-stage with court-by-year fixed effects they are replaced in the regressions by court-by-year time varying covariates---namely, the average number of judges, the average pretrial release rate, and the number of prosecutions (within a court in a given year).}
}
\end{table}

\begin{table}[h]
\captionsetup{justification=centering,margin=0cm}
\RawCaption{\caption{Crime Type Distribution}\label{Tab_Crime_type}}
{}
{\small\begin{tabular}{lcccc}\toprule
 & \multicolumn{2}{c}{Chilean} & \multicolumn{2}{c}{Immigrant}   \\ 
 & \% & N & \% & N   \\ \cline{2-5}
 &  &  &  &  \\ [-3pt]
 Homicide                           &         0.01 &        6,384 &         0.01 &           69   \\ [3pt]
 
 Sexual offense                                &         0.02 &       12,275 &         0.03 &          188   \\ [3pt]
 
 Theft or robbery                            &         0.26 &      180,086 &         0.17 &        1,104   \\ [3pt]
 
 Other property crime                    &         0.18 &      126,335 &         0.16 &        1,038   \\ [3pt]
 
  Drug offense                                &         0.12 &       85,936 &         0.21 &        1,306   \\ [3pt]

White-collar or tax crime                      &         0.02 &       11,946 &         0.02 &          142   \\ [3pt]

Crime against public trust                &         0.06 &       44,031 &         0.07 &          464   \\ [3pt]

 Crime against people's freedom and privacy                       &         0.29 &      198,834 &         0.29 &        1,816   \\ [3pt]

 Other crimes                                 &         0.04 &       27,179 &         0.04 &          235   \\ [3pt]
\bottomrule
\end{tabular}
\captionsetup{justification=justified,margin=0cm}
\floatfoot{\scriptsize\textbf{Notes:} This table presents the crime type distribution, by nationality, for the estimation sample. Shares are calculated to sum 100\% within Chileans and immigrants.}
}
\end{table}
\clearpage

\section{Data appendix}
\label{data_app}

This appendix gives a more detailed description of the data, the sample restrictions, and the construction of the variables.

\subsection{Sources} We merge two different sources of data to build our database.

\paragraph{PDO administrative records} We use administrative records from the Public Defender Office (PDO, see \href{http://www.dpp.cl/}{http://www.dpp.cl/}). The PDO is a centralized public service under the oversight of the Ministry of Justice that provides criminal defense services to all individuals accused of or charged with a crime who lack an attorney. The centralized nature of the PDO ensures that the administrative records contain information for all the cases handled by the PDO alone or those handled in coordination with a private attorney (as opposed to cases handled only by a private attorney), which covers more than 95\% of the universe of criminal cases in Chile. The unit of analysis is a criminal prosecution and contains defendants characteristics (ID, name, gender, nationality, and place of residence, among other characteristics) and case characteristics (case ID, court, public attorney assigned, initial and end dates, different categories for the type of crime, pretrial detention status and length, and outcome of the case, among other administrative characteristics). We consider cases whose arraignment hearings occurred between 2008 and 2017.

\paragraph{Registry of judges} In addition, we have access to information on arraignment judges and their assigned cases for arraignment hearings that occurred between 2008 and 2017. We merge this registry with the administrative records using the cases' IDs. We do not observe other characteristics of the judges other than their names and IDs. This data was shared by the Department of Studies at the Chilean Supreme Court (\href{https://www.pjud.cl/corte-suprema}{https://www.pjud.cl/corte-suprema}).  

\subsection{Estimation sample} The initial sample contains $3,571,230$ cases and covers all the cases recorded by the PDO that had an arraignment hearing between 2008 and 2017. To create our estimation sample, we make the following adjustments.

\paragraph{Basic data cleaning} Due to potential miscoding, we drop observations where the initial date of the case is later than the end date, and we also drop observations where the length of pretrial detention is greater than the length of the case. After these adjustments, the sample size reduces to $3,559,019$ (i.e., the number of cases reduces by $12,211$).

\paragraph{Sample restrictions} We then make the following sample restrictions:

\begin{itemize}
    
        \item We exclude hearings due to legal summons ($1,233,909$ observations). We do this because the information set available to the judges is likely to be different.
    
    \item We drop cases involving juvenile defendants ($254,243$ observations). We do this because the juvenile criminal justice system works differently, and the mandated selection rule and the preventive measures differ between systems (see \citealp{Cortes_et_al2019} for details).
    
    \item We drop cases where the defendant hires a private attorney as their exclusive defender ($103,092$ observations). We do this because we do not observe the result of the arraignment hearing (and what happens after in the prosecution) in these cases. 
    
    \item We drop cases that are longer than two years in duration ($55,495$ observations).
    
    \item For defendants that are accused of more than one crime in a given case and the records provide multiple observations, we consider the most severe crime (see below for the severity definition). In this step we drop $200,412$ observations. To be clear, we do not drop defendants, only cases. We do this to have at most one case/defendant pair per day of arraignment hearing.
    
    \item We drop cases where the detention judge ID is missing ($66,975$ observations). 

      \item We drop the types of crime with a likelihood of pretrial detention that is less than 5\% ($942,677$ observations). We do this because we want to study the decisions of judges in cases where pretrial detention is a plausible outcome.
    
    \item We drop cases handled by judges that see less than 10 cases in the whole time period ($2,848$ observations). We also resolve to only consider cases in which the assigned public attorney has defended at least 10 cases previously. It was not necessary to drop any data because of this restriction.

\end{itemize}
After all these adjustments the sample size is $699,368$, which is consistent with the figure in Table \ref{Tab_Desc_est}.

\subsection{Variables} Many of the variables used in our estimations are directly contained in the administrative records. In what follows we describe how we construct the other variables.

\begin{itemize}
    
        \item Severity: we proxy crime severity by computing the share of cases within the type of crime in which the defendants are detained pretrial.   
    \item Criminal record: we can track all the arrests of a given defendant using their IDs. Then, the variables previous prosecution, number of previous prosecutions, previous pretrial misconduct, previous conviction, and severity of previous prosecution are constructed by looking at the characteristics of the cases associated to the defendant's identification ID that were initiated before the current case. For individuals with no previous prosecutions, these variables are set to zero. To build these variables, we can track cases from 2005 onwards.
    
    \item Pretrial misconduct: pretrial misconduct is an indicator variable that takes value one if the defendant does not return to a scheduled hearing or is engaged in pretrial recidivism, or both. Nonappearance in court is recorded in the administrative data. Pretrial recidivism is built by looking at the arrests associated to the particular defendant's ID with an initial date that is between the initial and end dates of the current prosecution.
    
    \item Attorney quality and judge leniency: as in \cite{dgy}, we use the residualized (against court-by-time fixed effects) leave-out mean release rate. 
    \item Court-by-year of prosecution fixed effects: we consider the initial date to set the fixed effects.
\end{itemize}

\paragraph{Crime categories} We classify crimes following the PDO classification and group them in the following nine categories.
\begin{itemize}
    \item Homicides: considers all homicides, including specific categories such as parricide and femicide, among other specific types. 
    \item Sexual offenses: examples include sexual abuse, pedophilia, and rape, among other sex crimes. 
    \item Thefts and robberies: includes robbery, burglary, theft, and larceny.
    \item Other property crimes: examples include receiving or possession of stolen goods, arson, and criminal damages. 
    \item Drug offenses: includes illegal consumption, drug trafficking, and drug production. 
    \item White-collar and tax crimes: examples include economic fraud and the falsification of money, checks, or credit cards. 
    
    \item Crimes against public trust: examples include falsification of public, official, and commercial documents, forgery of private documents, falsification of certificates, and identity theft. 
    \item Crimes against the freedom and privacy of people: considers threats against citizens, but also includes threats to police officers and trespassing. 
    
    \item Other crimes: examples include gun possession and intellectual property theft.
\end{itemize}

\newpage
\section{Kitawaga-Oaxaca-Blinder decompositions} \label{KOB}

\setcounter{equation}{0} \renewcommand{\theequation}{C.\Roman{equation}}

\setcounter{table}{0} \renewcommand{\thetable}{C.\Roman{table}}

\setcounter{figure}{0} \renewcommand{\thefigure}{C.\Roman{figure}}

Let $\overline{R}_g = \mathbb{E}[Release_i|I_i=g]$, with $g\in\{0,1\}$. With KOB decompositions, group differences, $\Delta_R = \overline{R}_1 - \overline{R}_0$, can be explained using a vector of observed covariates, $X_i$. To do this, we first run an OLS projection for each group, $Release_i = X_i'\beta_g^R + \epsilon_{ig}^R$, where $X_i$ includes a constant and the individual controls of the benchmark equation, and $\epsilon_{ig}^R$ is the OLS projection error. By construction, OLS fits group means, so $\overline{R}_g = \overline{X}_g'\beta_g^R$, with $\overline{X}_g = \mathbb{E}[X_i|I_i=g]$. Then
\begin{eqnarray}
\Delta_R = \overline{X}_1'\beta_1^R - \overline{X}_0'\beta_0^R = \left(\overline{X}_1 - \overline{X}_0\right)'\beta_1^R + \overline{X}_0'\left(\beta_1^R - \beta_0^R\right).
\end{eqnarray}
The first term accounts for differences in release rates given differences in observables. The second term accounts for differences in release rates between defendants with the same observables (i.e., for differences in the estimated coefficients). When $X_i$ includes all the relevant characteristics that matter for the release decision, the second term can be interpreted as discrimination. If there are unobserved variables that correlate with $I_i$ and matter for misconduct potential, however, the OLS coefficients will capture their effect and the latter term will mistakenly interpreted as discrimination because of omitted variable bias (OVB). 

Our intuition is that if differences in unobservables matter for the release decision in a nondiscriminatory fashion, then they should be relevant to explain differences in pretrial misconduct rates when released. Formally, let $\overline{PM}_g = \mathbb{E}[PM_i|I_i=g,Release_i=1]$, where $PM_i$ is an indicator that takes value one if defendant $i$ engages in pretrial misconduct. Using the same logic as before, we can estimate $PM_i = X_i'\beta_g^P + \epsilon_{ig}^P$ in the sample of released defendants and write 
\begin{eqnarray}
\Delta_{PM} = \overline{X}_1'\beta_1^P - \overline{X}_0'\beta_0^P = \left(\overline{X}_1 - \overline{X}_0\right)'\beta_1^P + \overline{X}_0'\left(\beta_1^P - \beta_0^P\right).
\end{eqnarray}
Then, one way of testing if unobservables are important for interpreting our results is checking whether differences in observables are capable of explaining differences in pretrial misconduct rates. If the second component is large in the release equation but small in the outcome equation, then we can conjecture that unobservables are only playing a small role in explaining release rate disparities in a statistical sense, and we can therefore confidently interpret the benchmark estimations as evidence of discrimination.

\begin{table}[h]
\captionsetup{justification=centering,margin=0cm}
\RawCaption{\caption{Kitawaga-Oaxaca-Blinder Decomposition}\label{Tab_Oaxaca}}
{}
{\small\begin{tabular}{lcccc}\toprule
 & \multicolumn{2}{c}{Release} &  \multicolumn{2}{c}{Pretrial Misconduct}  \\  
 & Raw & Residualized &  Raw & Residualized  \\  
 & (1) & (2) & (3) & (4)   \\ \cline{2-5}
 &  &  &  &   \\ [-1pt]
 Total Difference                 &        0.067     &        0.030     &        0.050     &        0.076     \\ [1pt]
                                  & (0.011)  & (0.011)  & (0.009)  & (0.007)  \\ [5pt]
 Explained: Due to difference     &       -0.036    &       -0.056    &        0.065    &        0.076    \\ [1pt]
 in characteristics               & (0.003) & (0.003) & (0.003) & (0.004) \\ [5pt]
 Unexplained: Due to differences  &        0.103    &        0.085    &       -0.014    &       -0.001    \\ [1pt]
 in coefficients                  & (0.011) & (0.010) & (0.008) & (0.004) \\ [6pt]
 No. of Immigrants  &        6,362 &        6,362 &        4,900 &        4,900 \\ [2pt]
 No. of Chileans    &      693,006 &      693,006 &      580,400 &      580,400 \\ [5pt]
\bottomrule
\end{tabular}
\captionsetup{justification=justified,margin=0cm}
\floatfoot{\footnotesize\textbf{Notes:} This table presents the Kitawaga-Oaxaca-Blinder decomposition for release and pretrial misconduct estimation, considering raw and residualized covariates (residualized against court-by-year fixed effects).}
}
\end{table}

Table \ref{Tab_Oaxaca} shows the results. For each dependent variable, we present two versions of the KOB decompositions. Columns labeled as \textit{raw} present the standard KOB decomposition using the individual controls of the benchmark regressions as the vector of observables. Columns labeled as \textit{residualized} include dependent variables and the vector of individual controls residualized against court-by-year fixed effects. The release equation (columns 1 and 2) shows that there is a large share of the variation in release rates that cannot be explained by observed characteristics. In absolute value, the unexplained component of the average release gap is between two and three times larger than the share of variation explained by observables. Moreover, and consistent with the analysis so far, both components shift unconditional release disparities in opposite directions. In the absence of relevant unobserved variables, this reinforces the hypothesis of discrimination suggested by the benchmark regressions; however, it could also reflect the presence of OVB.

Columns 3 and 4 replicate the analysis using observed pretrial misconduct rates (among released defendants) as dependent variable. Note that in this case the complete average difference in pretrial misconduct rates can be explained by the same observed characteristics. This suggests that there are no relevant unobservables that explain misconduct potential, which reinforces the idea that the main benchmark regressions are not affected by OVB. Moreover, because observables do a good job of explaining actual misconduct rates, it reinforces the idea that our specific vector of observables makes the analysis closer in spirit to the disparate impact perspective, in the sense that this particular set of variables seem to do a reasonably good job of explaining observed misconduct rates. Or, at least, relevant omitted variables do not seem to be correlated with immigration status.

\newpage

\section{Outcome test}
\label{OT}

\setcounter{equation}{0} \renewcommand{\theequation}{D.\Roman{equation}}

\setcounter{table}{0} \renewcommand{\thetable}{D.\Roman{table}}

\setcounter{figure}{0} \renewcommand{\thefigure}{D.\Roman{figure}}

This appendix describes the outcome test, specifically the observational implementation proposed by \cite{pbot}, and provides suggestive evidence that the proposed identification argument is valid in our setting. 

\paragraph{Outcome test} The outcome test identifies a combination of biased beliefs and taste-based discrimination using observed outcomes of marginally selected individuals. Formal proofs are provided in \cite{ady}, \cite{pbot}, and \cite{hull2021}. In what follows, the intuition for the outcome test is presented.

Recall in Section \ref{sec2} that the release decision can be conceptualized as follows:
\begin{eqnarray}
Release_{i} = 1\left\{p(I_i,Z_i) \leq t_{j(i)}(I_i,Z_i) - b_{j(i)}(I_i,Z_i)\right\}. \label{ap_c1}
\end{eqnarray}
The outcome test examines whether the effective thresholds, $t_{j(i)}(I_i,Z_i) - b_{j(i)}(I_i,Z_i)$, are systematically different between immigrant defendants and Chilean defendants. Put formally, it tests whether
\begin{eqnarray}
\mathbb{E}\left[t_{j(i)}(1,Z_i) - b_{j(i)}(1,Z_i)\right]-\mathbb{E}\left[t_{j(i)}(0,Z_i) - b_{j(i)}(0,Z_i)\right] 
\end{eqnarray}
is different from zero, where the expectation is taken across $Z_i$ and $j(i)$. If the difference is not zero, then defendants with equal ``true'' pretrial misconduct probabilities will be detained at different rates. Notably, if a group is, on average, more prone to be engaged in pretrial misconduct, this does not affect the results of the outcome test. Differences in pretrial misconduct potential will affect the ``LHS'' of \eqref{ap_c1}; the OT estimates differences in the ``RHS''. If a group is systematically more risky, then it will cross the threshold more often, which is different from having a different threshold. That is why the outcome test identifies a notion of discrimination that abstracts from accurate sources of statistical discrimination. If judges are engaged in accurate statistical discrimination, then the outcome test should not be rejected \citep{hull2021}.

The insight produced by the outcome test is that although $t_{j(i)}(I_i,Z_i) - b_{j(i)}(I_i,Z_i)$ is not observed, for defendants that were released on a borderline decision $p(I_i,Z_i) = t_{j(i)}(I_i,Z_i) - b_{j(i)}(I_i,Z_i)$, and therefore the average misconduct rates of marginally released defendants identify $t_{j(i)}(I_i,Z_i) - b_{j(i)}(I_i,Z_i)$. Then, the outcome test is reduced to a difference in means that tests whether misconduct rates are different between marginally released immigrant defendants and marginally released Chilean defendants. 

The main identification challenge, then, is to identify marginal individuals. \cite{ady} provide a quasi-experimental approach that relies on the quasi-random assignment of judges. That approach is not applicable in our setting given the small share of immigrant defendants (the required instrument is underpowered). \cite{pbot} propose an observational approach that does not require instruments, which is the one implemented in this paper.

\paragraph{P-BOT} The prediction-based outcome test (P-BOT) proposed by \cite{pbot} uses the propensity score to identify marginal individuals. More specifically, \cite{pbot} provide sufficient conditions under which released defendants with lower propensity scores are more likely to be marginal given their observables. The implementation of the outcome test then proceeds as follows. First, we estimate the propensity score and compute the predicted values. Second, we rank released defendants according to their predicted release probabilities and define as marginal the released defendants at the bottom of the distribution. Third, we implement differences in means for pretrial misconduct rates between immigrant defendants and Chilean defendants who were marginally released.

Identification requires three assumptions and here we present a high-level discussion of these sufficient conditions (for technical details see \cite{pbot}). First, we need a common support assumption on the distribution of latent risk that allows us to claim that ``the more marginals'' are effectively marginals.  Second, the result relies on a separability assumption between observables and unobservables (by the econometrician) in the release equation. This implies that the effect of observables on the likelihood of being released is not affected by unobservables. Because the release decision is based on pretrial misconduct probabilities, this also implies a similar pattern in the outcome equation. Third, the result allows for unrestricted correlation between observables and unobservables but puts restrictions on the patterns of heteroskedasticity.

In what follows we present suggestive evidence that these assumptions are met in this setting implementing the tests as set out in \cite{pbot}.

\paragraph{P-BOT implementation} For implementing the test, we estimate the propensity score using the same set of observables included in the main benchmark estimations: the immigrant indicator, individual controls, judge and attorney controls, and court-by-time fixed effects. Then, we rank released defendants according to the predicted values and define the bottom 10\% of the distribution as marginals. Standard errors are bootstrapped considering that the sample selection rule is based on an estimated value. 

\begin{figure}[h]
\centering
\vspace{-5pt}
\caption{Immigrant Released Defendant and Chilean Released Defendant Propensity Score Histograms (Zoom up to 20th Percentile)}\label{Fig_hist}
\vspace{15pt}
\includegraphics[width=4in]{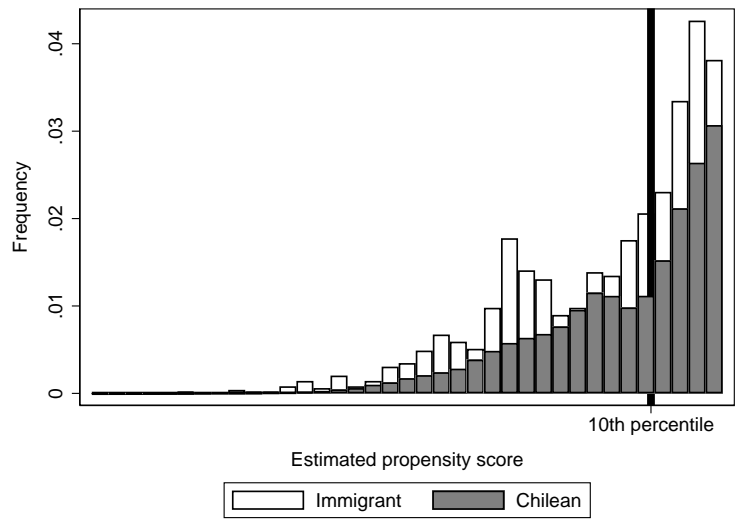}
\vspace{-10pt}
\floatfoot{\scriptsize\textbf{Notes:} This plot shows the propensity score histograms for immigrant released defendants and Chilean released defendants. The vertical line represents the 10th percentile of the distribution. For presentation purposes, we only show each histogram up to the 20th percentile. However, the histogram is calculated considering the entire population of released defendants.}
\end{figure}

\paragraph{Common support} Figure \ref{Fig_hist} shows the (estimated) propensity score distributions for released defendants, separating immigrant defendants and Chilean defendants. The figure suggests that the continuity and full support assumptions are met in our setting.

\paragraph{Monotonicity} One way to assess this assumption is to check whether the coefficients of a regression of $Release_i$ on covariates are stable (in terms of sign) when considering subsamples with (probably) different unobservables. Likewise, monotonicity also implies that the coefficients of a regression of pretrial misconduct on covariates are stable (in terms of sign) when considering subsamples with (probably) different unobservables.

Tables \ref{ap_monot_r} and \ref{ap_monot_pm} show the results using $Release_i$ and $PM_i$ as dependent variables, respectively. Each cell reports the estimated coefficient of the regressor specified in the column, using the sample specified in the first column. Each row represents a different estimation. The first row reports the coefficients using the whole sample. Thereafter, rows are paired by mutually exclusive sample categories that are (probably) characterized by different unobservables. For example, row 2 shows the results for the immigrant subsample, and row 3 shows the results for the Chilean subsample. Then, rows 4 and 5 split the sample by gender, and so on. The results strongly support the monotonicity assumption. In all but eight cases  (i.e., 90\% of cases) the sign of the coefficient is consistent across samples. Moreover, the magnitudes are also similar. This suggests that the direction of the effect of observables is unlikely to be affected by the unobserved variables.

\begin{table}[h!]
\captionsetup{justification=centering,margin=0cm}
\RawCaption{\caption{Testing for Monotonicity in Observables (Dep. Variable: Release Status)}\label{ap_monot_r}}

{}
{\small
\begin{tabular}{lccccc}
\toprule
& Previous & Previous pretrial  & Previous   & Severity       & Severity \\ 
 & case     & misconduct &   conviction & previous case  & current case \\ [-4pt]
 \textit{Estimation sample} &    &  &  &    &  \\ [5pt]
 All             &       -0.004  &       -0.026  &       -0.013 &       -0.136 &       -1.009  \\ 
                 & (0.003) & (0.001) & (0.003) & (0.004) & (0.003)  \\ [2pt] 
 Immigrant       &        0.044 &       -0.027 &        0.061 &       -0.129 &       -1.082  \\ 
                 & (0.033) & (0.017) & (0.032) & (0.058) & (0.034)  \\ [2pt] 
 Chilean         &       -0.006 &       -0.026 &       -0.014 &       -0.136 &       -1.007  \\ 
                 & (0.003) & (0.001) & (0.003) & (0.004) & (0.003)  \\ [2pt] 
 Male            &       -0.004 &       -0.027 &       -0.012 &       -0.119 &       -1.014  \\ 
                 & (0.003) & (0.001) & (0.003) & (0.004) & (0.003)  \\ [2pt] 
 Female          &        0.011 &       -0.010 &       -0.030 &       -0.341 &       -0.958  \\ 
                 & (0.007) & (0.003) & (0.007) & (0.013) & (0.009)  \\ [2pt] 
 Low income      &       -0.006 &       -0.021 &       -0.015 &       -0.131 &       -1.019  \\ 
                 & (0.004) & (0.002) & (0.004) & (0.006) & (0.005)  \\ [2pt] 
 High income     &       -0.003 &       -0.029 &       -0.012 &       -0.140 &       -1.002  \\ 
                 & (0.003) & (0.001) & (0.003) & (0.005) & (0.004)  \\ [2pt] 
 Low judge       &       -0.002 &       -0.028 &       -0.016 &       -0.151 &       -1.050  \\ 
 leniency        & (0.004) & (0.002) & (0.004) & (0.005) & (0.004)  \\ [2pt] 
 High judge      &       -0.005 &       -0.023 &       -0.011 &       -0.121 &       -0.968  \\ 
 leniency        & (0.004) & (0.002) & (0.003) & (0.005) & (0.004)  \\ [2pt] 
 Low attorney    &       -0.004 &       -0.025 &       -0.016 &       -0.144 &       -1.073  \\ 
 quality         & (0.004) & (0.002) & (0.004) & (0.005) & (0.004)  \\ [2pt] 
 High attorney   &       -0.005 &       -0.026 &       -0.010 &       -0.127 &       -0.940  \\ 
 quality         & (0.004) & (0.002) & (0.003) & (0.005) & (0.004)  \\ [2pt] 
 Small court     &        0.002 &       -0.022 &       -0.017 &       -0.153 &       -1.084  \\ 
 (No. of cases)  & (0.004) & (0.002) & (0.004) & (0.005) & (0.004)  \\ [2pt] 
 Big court       &       -0.009 &       -0.028 &       -0.011 &       -0.126 &       -0.943  \\ 
 (No. of cases)  & (0.004) & (0.002) & (0.003) & (0.005) & (0.004)  \\ [2pt] 
 Small court     &        0.003 &       -0.022 &       -0.017 &       -0.149 &       -1.077  \\ 
 (No. of judges) & (0.004) & (0.002) & (0.004) & (0.005) & (0.004)  \\ [2pt] 
 Big court       &       -0.009 &       -0.028 &       -0.012 &       -0.127 &       -0.946  \\ 
 (No. of judges) & (0.004) & (0.002) & (0.003) & (0.005) & (0.004)  \\ [2pt] 
 Low severity    &       -0.005 &       -0.019 &       -0.010 &       -0.111 &       -0.879  \\ 
 court           & (0.003) & (0.001) & (0.003) & (0.005) & (0.004)  \\ [2pt] 
 High severity   &       -0.002 &       -0.028 &       -0.019 &       -0.158 &       -1.132  \\ 
 court           & (0.004) & (0.002) & (0.004) & (0.005) & (0.004)  \\ [5pt] 
 
\bottomrule
\end{tabular}
\captionsetup{justification=justified,margin=0cm}
\floatfoot{\footnotesize\textbf{Notes:} This table presents the results of the test for monotonicity in observables. Each reported value is the marginal effect of the variable of the column on the probability of release, estimated using a different sample in each row. The continuous variables were discretized using the respective median as the threshold. The values in parentheses are standard errors.}
}
\end{table}

\begin{table}[h!]
\captionsetup{justification=centering,margin=0cm}
\RawCaption{\caption{Testing for Monotonicity in Observables (Dep. Variable: Pretrial Misconduct)}\label{ap_monot_pm}}
{}
{\small
\begin{tabular}{lccccc}
\toprule
& Previous & Previous pretrial  & Previous   & Severity       & Severity \\ 
 & case     & misconduct &   conviction & previous case  & current case \\ [-4pt]
 \textit{Estimation sample} &    &  &  &    &  \\ [5pt]
 All             &        0.050  &        0.106  &        0.036 &        0.036 &        0.034  \\ 
                 & (0.004) & (0.002) & (0.003) & (0.005) & (0.005)  \\ [2pt] 
 Immigrant       &        0.041 &        0.096 &        0.044 &        0.076 &        0.111  \\ 
                 & (0.040) & (0.020) & (0.039) & (0.073) & (0.045)  \\ [2pt] 
 Chilean         &        0.050 &        0.106 &        0.036 &        0.036 &        0.033  \\ 
                 & (0.004) & (0.002) & (0.004) & (0.005) & (0.005)  \\ [2pt] 
 Male            &        0.051 &        0.107 &        0.034 &        0.041 &        0.039  \\ 
                 & (0.004) & (0.002) & (0.004) & (0.006) & (0.005)  \\ [2pt] 
 Female          &        0.049 &        0.096 &        0.046 &       -0.024 &       -0.012  \\ 
                 & (0.010) & (0.005) & (0.010) & (0.020) & (0.014)  \\ [2pt] 
 Low income      &        0.045 &        0.097 &        0.038 &        0.036 &        0.075  \\ 
                 & (0.006) & (0.003) & (0.006) & (0.008) & (0.007)  \\ [2pt] 
 High income     &        0.051 &        0.112 &        0.035 &        0.037 &        0.000  \\ 
                 & (0.005) & (0.002) & (0.004) & (0.007) & (0.006)  \\ [2pt] 
 Low judge       &        0.041 &        0.102 &        0.044 &        0.043 &        0.035  \\ 
 leniency        & (0.005) & (0.002) & (0.005) & (0.008) & (0.007)  \\ [2pt] 
 High judge      &        0.059 &        0.110 &        0.027 &        0.029 &        0.034  \\ 
 leniency        & (0.005) & (0.002) & (0.005) & (0.008) & (0.007)  \\ [2pt] 
 Low attorney    &        0.047 &        0.110 &        0.042 &        0.048 &        0.033  \\ 
 quality         & (0.005) & (0.002) & (0.005) & (0.008) & (0.007)  \\ [2pt] 
 High attorney   &        0.054 &        0.102 &        0.029 &        0.023 &        0.034  \\ 
 quality         & (0.005) & (0.002) & (0.005) & (0.008) & (0.007)  \\ [2pt] 
 Small court     &        0.039 &        0.102 &        0.035 &        0.043 &        0.094  \\ 
 (No. of cases)  & (0.005) & (0.002) & (0.005) & (0.008) & (0.007)  \\ [2pt] 
 Big court       &        0.061 &        0.107 &        0.038 &        0.025 &       -0.024  \\ 
 (No. of cases)  & (0.005) & (0.002) & (0.005) & (0.007) & (0.006)  \\ [2pt] 
 Small court     &        0.054 &        0.103 &        0.030 &        0.040 &        0.057  \\ 
 (No. of judges) & (0.005) & (0.002) & (0.005) & (0.008) & (0.007)  \\ [2pt] 
 Big court       &        0.047 &        0.107 &        0.042 &        0.030 &        0.013  \\ 
 (No. of judges) & (0.005) & (0.002) & (0.005) & (0.007) & (       0.006)  \\ [2pt] 
 Low severity    &        0.045 &        0.101 &        0.039 &        0.042 &        0.053  \\ 
 court           & (0.005) & (0.002) & (0.005) & (0.007) & (0.006)  \\ [2pt] 
 High severity   &        0.053 &        0.108 &        0.034 &        0.028 &        0.016  \\ 
 court           & (0.005) & (0.002) & (0.005) & (0.008) & (0.007)  \\ [5pt] 
 
\bottomrule
\end{tabular}
\captionsetup{justification=justified,margin=0cm}
\floatfoot{\footnotesize\textbf{Notes:} This table presents the results of the test for monotonicity in observables. Each reported value is the marginal effect of the variable of the column on pretrial misconduct, estimated using a different sample of released defendants in each row. The continuous variables were discretized using the respective median as the threshold. The values in parentheses are standard errors.}
}
\end{table}

\paragraph{Ranking validity} Assume that our set of observed variables, $X_i$, is a good approximation (up to some small well-behaved noise) of the (complete) information set of a judge. Under this assumption, the identification of marginally released defendants using the ranking based on the propensity score is accurate. We fit the propensity score and label as marginal the bottom 10\% of the predicted probability distribution (among released defendants ). Then, we omit one observable and (i) estimate the propensity score with the restricted set of observables and identify marginals using the ranking strategy, and (ii) compute the conditional probabilities of being marginal---namely, the shares of marginals identified in the first step for different combinations of the observables used in the restricted estimation. We then compute the rank correlation between (i) the share of marginals using the restricted propensity-score ranking and the conditional probabilities, and (ii) the estimated propensity score using the restricted set of observables and the conditional probabilities of being marginal. In case (i), the correlation is expected to be positive. In case (ii), the correlation is expected to be negative. If the identification argument holds, we should expect these rank correlations to be large.

We perform this exercise by using each of the 14 observables used in the estimation.\footnote{The variables are number of previous cases, severity of previous case, severity of current case, average severity by year-court, number of cases by year-court, judge leniency, judge leniency squared, attorney quality, attorney quality squared, immigrant indicator, previous case indicator, previous pretrial misconduct indicator, and previous conviction indicator.} To compute the rank correlations, we discretize the nondiscrete regressors (using the median) to define $2^{(14-1)}=8,192$ categories of observables. For each of these categories, we compute the average restricted estimated propensity score, the average share of marginals using the restricted propensity score, and the conditional probability of being marginal using the base estimation as the true share of marginals. Table \ref{tab_rank} presents the results. We report both Spearman's-$\rho$ and Kendall's-$\tau$ statistics for rank correlation. In all variables bar one (severity of current case) the correlations are very large. We interpret this as strong suggestive evidence of the validity of the identification argument. 

\begin{landscape}
\begin{table}[h]
\captionsetup{justification=centering,margin=0cm}
\RawCaption{\caption{Rank Correlations}\label{tab_rank}}
\floatbox[{\captop}]{table}[\FBwidth]
{}
{\small
\begin{tabular}{lcccc}\toprule
&  &  &  &  \\ [-7pt]
& \multicolumn{2}{c}{Corr. btw. $\Pr(Marg| X=x,Release=1)$} & \multicolumn{2}{c}{Corr. btw. $\Pr(Marg| X=x,Release=1)$}  \\ 
 & \multicolumn{2}{c}{and $\mathbb{E}[Marg| X=x]$ using restricted p-score} & \multicolumn{2}{c}{and $\mathbb{E}[Release| X=x]$ using restricted p-score} \\ [8pt] 
\textit{ Excluded predictor} & Spearman & Kendall & Spearman & Kendall \\
 &  &  & &  \\ [-5pt]
No of previous cases&       0.972&       0.948&      -0.668&      -0.549\\
Severity previous case&       0.980&       0.960&      -0.672&      -0.555\\
Severity current case&       0.434&       0.379&      -0.264&      -0.217\\
Average severity (year/court)&       0.960&       0.927&      -0.671&      -0.557\\
No of cases (year/court)&       0.997&       0.990&      -0.666&      -0.551\\
No of judges (year/court)&       0.986&       0.974&      -0.675&      -0.559\\
Judge leniency&       0.988&       0.976&      -0.673&      -0.556\\
Judge leniency square&       1.000&       0.999&      -0.678&      -0.560\\
Attorney quality&       0.975&       0.955&      -0.676&      -0.559\\
Attorney quality square&       1.000&       1.000&      -0.675&      -0.558\\
Immigrant   &       0.997&       0.992&      -0.806&      -0.661\\
Previous case&       0.998&       0.996&      -0.666&      -0.551\\
Previous pretrial misconduct&       0.989&       0.981&      -0.678&      -0.562\\
Previous conviction&       1.000&       0.998&      -0.670&      -0.556\\
 
\bottomrule
\end{tabular}
\captionsetup{justification=justified,margin=0cm}
\floatfoot{\footnotesize\textbf{Notes:} This table presents the rank correlations between the ranking of the conditional probabilities of being marginal and (i) the ranking of the conditional share of marginals  using the restricted propensity score estimation, and (ii) the ranking of the predicted propensity score using the restricted estimation. We report Spearman's-$\rho$ and Kendall's-$\tau_b$ rank correlation statistics. The excluded predictor is specified in the first column. All regressions include year fixed effects. The unit of analysis to build the ranking is the combination of all possible values of the predictors, without considering the excluded category (i.e., 13 predictors). The continuous predictors were transformed into binary variables using the median among released defendants  as the threshold.}
}
\end{table}
\end{landscape}

\end{document}